\documentclass[bibyear]{aa} % if the references are not structured 
\usepackage{graphicx}
\usepackage{txfonts}
\begin{document}

   \title{The role of planetesimal fragmentation on giant planet formation}

   \author{O. M. Guilera$^{1,2}$; G. C. de El\'{\i}a$^{1,2}$; A. Brunini$^{1,2}$ \and P. J. Santamar\'{\i}a$^{1,2}$}

   \institute{Grupo de Ciencias Planetarias, Facultad de Ciencias Astron\'omicas y Geof\'{\i}sicas, Universidad Nacional de La Plata, Argentina.    
             \and
             Grupo de Ciencias Planetarias,  Instituto de Astrof\'{\i}sica de La Plata (Consejo Nacional de Investigaciones Cient\'{\i}ficas y T\'ecnicas - Universidad Nacional de La Plata), Argentina.}

   \offprints{oguilera@fcaglp.unlp.edu.ar}
   
   \date{Received }

% \abstract{}{}{}{}{} 
% 5 {} token are mandatory
 
  \abstract
  {In the standard scenario of planet formation, terrestrial planets and the 
   cores of the giant planets are formed by accretion of planetesimals. As 
   planetary embryos grow the planetesimal velocity dispersion
   increases due to gravitational excitations produced by embryos. The  
   increase of planetesimal relative velocities causes the fragmentation
   of them due to mutual collisions.} 
  {We study the role of planetesimal fragmentation on giant planet
  formation. We analyze how planetesimal fragmentation modifies the
  growth of giant planet's cores for a wide range of planetesimal
  sizes and disk masses.} 
  {We incorporate a model of planetesimal fragmentation into our model of in 
   situ giant planet formation. We calculate the evolution of the solid
   surface density (planetesimals plus fragments) due to the accretion by 
   the planet, migration and fragmentation.} 
   {The incorporation of planetesimal fragmentation significantly
   modifies the process of planetary formation. If most of the mass loss
   in planetesimal collisions is distributed in the smaller fragments,
   planetesimal fragmentation inhibits the growth of the embryo for
   initial planetesimals of radii lower than 10~km. Only for initial
   planetesimals of 100~km of radius, and disks greater than
   $0.06~\textrm{M}_{\odot}$, embryos achieve masses greater than the
   mass of the Earth. However, even for such big planetesimals and
   massive disks, planetesimal fragmentation induces the quickly
   formation of massive cores only if most of the mass loss in
   planetesimal collisions is distributed in the bigger fragments.}   
   {Planetesimal fragmentation seems to play an important role in
   giant planet formation. The way in which the mass loss in planetesimal collisions is distributed leads to different results, inhibiting or favoring the
   formation of massive cores.}  


   \maketitle
%
%________________________________________________________________

\section{Introduction}

According to the core accretion model (Lissauer \& Stevenson,
2007) the formation of a giant planet occurs through a sequence of
events:  
\begin{itemize} 
\item initially, the dust (particles of $\sim \mu\textrm{m}$ sizes)
      collapses to the protoplanetary disk mid-plane,
\item by different mechanisms --still under discussion-- these particles
      agglomerate between them leading to the formation of planetesimals
      (objects between hundreds of meters and hundreds of kilometers),
\item planetesimals grow by mutual accretions until some bodies begin
      to differentiate themselves from the population of planetesimals
      (planetary embryos, objects with sizes of few thousand
      kilometers),
\item embryo gravitational excitation over planetesimals limits the
      growth of them, and embryos are the only bodies that grow by
      accretion of planetesimals,  
\item as embryos grow they bound a gaseous envelope. Initially, the
      planetesimal accretion rate is much higher than the gas accretion
      rate, 
\item when the mass of the envelope reaches a critical value of the
      order of the mass of the solid core, the envelope layers cannot
      remain in hydrostatic equilibrium and a gaseous runaway growth
      process starts, 
\item finally -- by mechanisms also still under discussion-- the planet
      stops the gas accretion and evolves contracting and colling at
      constant mass.  
\end{itemize}

Thus, it is important to study the evolution of the population of
planetesimals together with the process of accretion and planet
formation. The evolution of the population of planetesimals is a complex
phenomenon. Planetesimal accretion by the embryos, migration due to gas
drag produced by the gaseous component of the protoplanetary disk,
collisional evolution due to gravitational excitations produce by
embryos, planetesimal dispersion and gap openings, maybe are the most
relevant phenomena.

Regarding fragmentation, as embryos grow they increase planetesimal
 relative velocities causing planetesimal fragmentation. After successive
 disruption collisions --also called collisional cascade-- planetesimals
 become smaller. Inaba et al. (2003), Kobayashi et al. (2011) and Ormel
 \& Kobayashi (2012) found that a significant amount of mass, that
 remains in small fragments product of the collisions between
 planetesimals, may be lost by migration due to gas drag. So, the planetesimal fragmentation seems to play an important role in forming the cores of giant
 planets. Moreover, as embryos grow, they begin to bind the surrounding
 gas. Initially, embryo gaseous envelopes are less massive but
 wide-spreads. These envelopes produce a loss of the planetesimal
 kinetic energy, significantly increasing the capture cross section of
 the planets. The smaller planetesimals of the distribution more suffer
 both effects. So, while smaller fragments have higher migration rates
 due to gas drag, they are more efficiently accreted by the
 planet. There is a strong competition between the time scales of
 migration and accretion of small fragments generated by planetesimal
 fragmentation. Therefore, it is important to study in detail if the
 generation of small fragments --products of planetesimal
 fragmentation-- favors or inhibits giant planet formation. 

In previous works (Guilera et al., 2010; 2011) we developed a model for
giant planet formation, which calculates the formation of them immersed
in a protoplanetary disk that evolves in time. In these previous works
the population of planetesimals evolved by the accretion of the embryos and by
planetesimal migration. In this new work, we incorporated the
fragmentation of planetesimals to study if this phenomenon produces
significant changes --and thus if it is a primary key to consider-- in the
process of planetary formation.

This work is organized as follows: in Section 2 we introduce some improvements to our previous model; in Section 3, we explain in detail the
planetesimal fragmentation model adopted; Section 4 shows the results
of the role of planetesimal fragmentation on the growth of an embryo
located at 5~AU; finally, in Section 5 we present the conclusions about the
results found in this work.    

%__________________________________________________________________

\section{Improvements to our previous model}

Our model describing the evolution of the protoplanetary disk  is
based on the works of Guilera et al. (2010; 2011) with some minor
improves incorporated. We used an axisymmetric protoplanetary disk
characterized by a gaseous and solid component. The gaseous component is 
represented by a 1D grid for the radial coordinate, while the solid
component is represented by a 2D grid, where one dimension is for the
radial coordinate and the other one is for the different planetesimal
sizes. Some quantities are only functions of the radial coordinate ($R$)
--like the gas surface density $\Sigma_g(R)$-- while some others are
also functions of the planetesimal sizes, like the planetesimal surface
density $\Sigma_p(R,r_p)$. 

\subsection{Planetesimal size distribution}

We change the treatment of the planetesimal size distribution respect
to ours previous work. We now consider that the different radii are
logarithmic equally spaced, so the $j$ species is given by, 
\begin{eqnarray}
r_{p_j}= \left( \frac{r_{p_N}}{r_{p_1}} \right)^{\frac{j-1}{N-1}} r_{p_1}, \quad j= 1,...,N,
\label{eq:size-step}
\end{eqnarray}  
where $r_{p_N}$ and $r_{p_1}$ are the maximum and minimum radii of the
size distribution, respectively, and $N$ is the number of size bins
considered. If $m_{p_j}$ is the mass between $m_{p_{j-1/2}}$ and
$m_{p_{j+1/2}}$, and adopting that the planetesimal mass distribution
is represented by a power law ($dn/dm \propto m_p^{-\alpha}$), $m_{p_j}$
is given by,  
\begin{eqnarray}
m_{p_j} & = & \int_{m_{p_{j-1/2}}}^{m_{p_{j+1/2}}}~mn(m)~dm \nonumber \\ 
       & = & \frac{\textrm{C}m_{p_1}^{2-\alpha}}{2-\alpha} \left[ \Delta^{\frac{3(j-1/2)(2-\alpha)}{N-1}} - \Delta^{\frac{3(j-3/2)(2-\alpha)}{N-1}} \right],
\end{eqnarray} 
where we use that $m_{p_j}= \Delta^{3(j-1)/(N-1)} m_{p_1}$ with $\Delta= r_{p_N}/r_{p_1}$. In the same way, the total mass is given by 
\begin{eqnarray}
m_T & = & \int_{m_{p_{1-1/2}}}^{m_{p_{N+1/2}}}~mn(m)~dm, \nonumber \\
       & = & \frac{\textrm{C}m_{p_1}^{2-\alpha}}{2-\alpha} \left[ \Delta^{\frac{3(N-1/2)(2-\alpha)}{N-1}} - \Delta^{\frac{3(2-\alpha)}{2(N-1)}} \right].
\end{eqnarray}   
The amount of mass (respect to the total mass) corresponding to the $j$
species is given by, 
\begin{eqnarray}
p_j & = & \frac{m_{p_j}}{m_T},  \nonumber \\
    & = & \left[ \frac{\Delta^{\frac{3(j-1/2)(2-\alpha)}{N-1}} - \Delta^{\frac{3(j-3/2)(2-\alpha)}{N-1}}}{\Delta^{\frac{3(N-1/2)(2-\alpha)}{N-1}} - \Delta^{\frac{3(2-\alpha)}{2(N-1)}}} \right].
\end{eqnarray}
Finally, the planetesimal surface density corresponding to the
planetesimals of radius $r_{p_j}$ is obtained by
multiplying $p_j$ and the total surface density of solids. Then, we treat
each planetesimal size independently. In this approach we can use an
only planetesimal size (for this case $p=1$) or a discrete numbers ($N$)
of bins to approximate the continuous planetesimal size distribution.

\subsection{Evolution of planetesimal eccentricities, inclinations and velocity migrations}

As in our previous works, we consider that the evolution of the
eccentricities and inclinations of the planetesimals are governed by two
mayor processes: the embryo gravitational excitations and the damping due to the gas drag. 

The embryo stirring rates of the eccentricities and inclinations are
given by (Ohtsuki et al., 2002), 
\begin{eqnarray}
\frac{de^2}{dt} \Big|_{stirr} & = &  \left(
				      \frac{\textrm{M}_{\textrm{P}}}{3bM_{\star}P_{orb}}
				     \right) P_{stirr}, \\ 
\frac{di^2}{dt} \Big|_{stirr} & = &  \left( \frac{\textrm{M}_{\textrm{P}}}{3bM_{\star}P_{orb}}
				     \right) Q_{stirr}, 
\end{eqnarray}
where $\textrm{M}_{\textrm{P}}$ is the mass of the planetary embryo, $M_{\star}$ is the mass
of the central star, $b$ is the full width of the feeding zone of the
planetary embryo in terms of its Hill radius and $P_{orb}$ is the orbital
period of the embryo. Finally, $P_{stirr}$ and $Q_{stirr}$ are functions
of the planetesimal eccentricities and inclinations (for further details
see Chambers, 2006). However, this is a local approach. The
gravitational excitation decreases with increasing distance between the
planetary embryo and the planetesimals. Hasegawa \& Nakasawa (1990)
showed that when 
the distance from the planetary embryo is larger than $\sim 4$ times its
Hill radius, the excitation over the planetesimals decays
significantly. Therefore, we need to restrict this effect to the
neighborhood of the planetary embryo. Using the EVORB code (Fernandez et
al., 2002) we fit a a modulation function in order to reproduce the excitation over the quadratic mean value of the eccentricity of a planetesimal. We find this excitation is well reproduced by, 
\begin{eqnarray}
f(\Delta) = \left( \frac{1}{1 + \left| \frac{\Delta}{2.85\textrm{R}_{\textrm{H}}} \right|^{10}} \right),
\label{eq:mod_fun}   
\end{eqnarray}
where $\Delta= R - R_P$ represents the distance from the planet ($R_P$
is the planet radius orbit), $\textrm{R}_{\textrm{H}}$ is the
planetary embryo Hill radius and $f(\Delta)$ guarantees that the
eccentricity and inclination profiles of the planetesimals are smooth
enough for a numerical treatment  and that the planetary excitation on
planetesimals is restricted to the embryo neighborhood. 

On the other hand, the eccentricities and inclinations of the
planetesimals are damped by the gaseous component of the
protoplanetary disk. This damping depends on the planetesimal relative
velocity with respect the gas, $v_{rel}^{p-g}$, and on the ratio between
planetesimal radius and the molecular mean free path, $\lambda$. Adopting a
gaseous disk mainly composed by molecular hydrogen ($\textrm{H}_2$),
the last is given by (Adachi, 1976), 
\begin{eqnarray}
\lambda_{\textrm{H}_2}= \frac{\mu_{\textrm{H}_2}}{\sqrt{2} \pi \rho_g
 d_{\textrm{H}_2}}, 
\end{eqnarray}
where $\mu_{\textrm{H}_2}$ and $d_{\textrm{H}_2}$ are the molecular weight
and molecular diameter of the molecular hydrogen, respectively, and
$\rho_g$ is the volumetric gas density.   

As in the recent work of Fortier et al. (2013), we consider three
different regimes (Rafikov, 2004; Chambers, 2008), 
\begin{itemize}
\item Epstein regime: $r_p < \lambda_{\textrm{H}_2}$,
\item Stokes regime: $r_p > \lambda_{\textrm{H}_2}$ and $Re <
      Re_{trans}$,
\item Quadratic regime: $r_p > \lambda_{\textrm{H}_2}$ and $Re >
      Re_{trans}$,
\end{itemize}
where $Re= v_{rel}^{p-g} r_p / \nu$ is the Reynolds number and
$Re_{trans}= 20$ is the transition between Stokes and Quadratic regimes
(Rafikov, 2004). The viscosity $\nu$ corresponds to the molecular
viscosity, given by,
\begin{eqnarray}
\nu= \frac{\lambda_{\textrm{H}_2} c_s}{3},
\end{eqnarray}
 where $c_s$ is the local speed of the sound.  

The incorporation of the different regimes is important due to the fact that smaller fragments (products of planetesimal fragmentation) could be in Stokes or Epstein regimes. The three drag regimes can be characterized in terms of the stopping time given by (Chambers, 2008),
\begin{eqnarray}
t_{stop}^{Eps} & = & \frac{\rho_p r_p}{\rho_g c_s}, \\ 
t_{stop}^{Sto} & = & \frac{2 \rho_p r_p}{3 \rho_g c_s \lambda_{\textrm{H}_2}}, \\
t_{stop}^{Qua} & = & \frac{6 \rho_p r_p}{\rho_g v_{rel}^{p-g}},      
\end{eqnarray} 
where $\rho_p$ is the planetesimal density. The relative velocity
between planetesimals and the gas is given by,
\begin{eqnarray}
v_{rel}^{p-g}= \sqrt{\eta^2 + \frac{5}{8}e^2 + \frac{1}{2}i^2}~v_k, 
\end{eqnarray}
where $v_k$ is the Keplerian velocity, $\eta= (v_k-v_g)/v_k$ is the
ratio of the gas velocity to the Keplerian velocity.

The damping rates of the eccentricities and inclinations for each regime
are given by (Rafikov, 2004; Chambers, 2008),
\begin{eqnarray} 
\frac{de^2}{dt} \Big|_{gas}^{Eps} & = & - \frac{2}{t_{stop}^{Eps}}
 \left(\frac{s^2_{Eps}}{1+s^2_{Eps}}\right) e^2, \\ 
\frac{di^2}{dt} \Big|_{gas}^{Eps} & = & - \frac{2}{t_{stop}^{Eps}}
 \left(\frac{s^2_{Eps}}{1+s^2_{Eps}}\right) i^2,      
\end{eqnarray} 
where $s_{Eps}= 2\pi t_{stop}^{Eps}/P_{orb}$, 
\begin{eqnarray} 
\frac{de^2}{dt} \Big|_{gas}^{Sto} & = & - \frac{2}{t_{stop}^{Sto}}
 \left(\frac{s^2_{Sto}}{1+s^2_{Sto}}\right) e^2, \\ 
\frac{di^2}{dt} \Big|_{gas}^{Sto} & = & - \frac{2}{t_{stop}^{Sto}}
 \left(\frac{s^2_{Sto}}{1+s^2_{Sto}}\right) i^2,      
\end{eqnarray} 
with $s^2_{Sto}= 2\pi t_{stop}^{Sto}/P_{orb}$, and
\begin{eqnarray} 
\frac{de^2}{dt} \Big|_{gas}^{Qua} & = & - \frac{2 e^2}{t_{stop}^{Qua}}, \\ 
\frac{di^2}{dt} \Big|_{gas}^{Qua} & = & - \frac{2 i^2}{t_{stop}^{Qua}}.
\end{eqnarray}
Finally, the evolution of the eccentricities and inclinations are given solving the coupled equations by,  
\begin{eqnarray}
\frac{de^2}{dt} & = & f(\Delta) \frac{de^2}{dt} \Big|_{stirr} + \frac{de^2}{dt}
 \Big|_{gas}, \\
\frac{di^2}{dt} & = & f(\Delta) \frac{di^2}{dt} \Big|_{stirr} + \frac{di^2}{dt}
 \Big|_{gas}.
\label{eq:evol_ei}  
\end{eqnarray}

The gas drag also causes an inward planetary orbit migration. Then, the
rate of change of the major semi-axis is give by, 
\begin{eqnarray} 
\frac{da}{dt}= v_{mig}= \left\{ 
\begin{array}{cc}
- \frac{2a\eta}{t_{stop}^{Eps}}\left(\frac{s^2_{Eps}}{1+s^2_{Eps}}\right)
 & \textrm{Epstein regime} \\
- \frac{2a\eta}{t_{stop}^{Sto}}\left(\frac{s^2_{Sto}}{1+s^2_{Sto}}\right)
 & \textrm{Stokes regime} \\
- \frac{2a\eta}{t_{stop}^{Qua}} & \textrm{Quadratic regime}
\end{array} \right.
\end{eqnarray}

\subsection{Oligarchic accretion regime}

As in previous works, we consider that the embryos grow in the oligarchic regime. Assuming the particle in a box approximation, the planetesimal accretion rate of the $j$ species is given by (Inaba et al., 2001), 
\begin{eqnarray}
\frac{d\textrm{M}_{\textrm{C}}^j}{dt}= \frac{2 \pi \Sigma_p(R_P,r_{p_j})r^2_{P_H}}{P_{orb}}P_{coll}(\textrm{R}_{\textrm{C}},\textrm{R}_{\textrm{H}}, v_{rel}^{p^j-P}), 
\label{eq:mc_acc}
\end{eqnarray}
where $\textrm{M}_{\textrm{C}}$ is the embryo's core mass and $P_{coll}$ is the collision
probability between the planetesimal $j$ species and the embryo (see
Guilera et al. 2010, for the explicit expression of $P_{coll}$). The collision
probability is a function of the embryo's core radius ($\textrm{R}_{\textrm{C}}$), the embryo's Hill radius, and the relative velocity between the planetesimal $j$ species and the embryo, which is given by,
\begin{eqnarray}
 v_{rel}^{p^j-P}= \sqrt{\frac{5}{8}e^2 + \frac{1}{2}i^2}~v_k.
\end{eqnarray}
When the embryo has a non negligible envelope we have to incorporate the
enhancement of capture cross section of the embryo. As in previous works
we use the prescription given by Inaba \& Ikoma (2003) to calculate the
embryo enhanced radius $\tilde{\textrm{R}}_{\textrm{C}}$, where they propose  replacing
$\tilde{\textrm{R}}_{\textrm{C}}$ for $\textrm{R}_{\textrm{C}}$ in the expressions of the collision probability. 

The feeding zone of the embryo is often defined as the ring around itself where
planetesimals can be accreted. We defined the width of the feeding zone
as $\sim 4$ times the embryo's Hill radius (at both sides of the embryo ). So, we integrate Eq.~({\ref{eq:mc_acc}}) over the radial grid 
\begin{eqnarray}
\frac{d\textrm{M}_{\textrm{C}}^j}{dt} = \frac{2 \pi r^2_{P_H}P_{coll}}{P_{orb}}~\int_{FZ} 2 \pi R \psi(R,R_P,\textrm{R}_{\textrm{H}})\Sigma_p(R,r_{p_j})~dR, 
\label{eq:mc_acc_int}
\end{eqnarray}
where $\psi(R,R_P,\textrm{R}_{\textrm{H}})$ is a normalization function which satisfies $\int_{-\infty}^{\infty} 2 \pi R \psi(R,R_P,\textrm{R}_{\textrm{H}})~dR = 1$. In contrast with our previous work we chose that 
\begin{eqnarray}
\psi= \frac{3e^{-\left(\frac{R-R_P}{4\textrm{R}_{\textrm{H}}}\right)^6}}{8\pi \textrm{R}_{\textrm{H}}R_P\Gamma(1/6)},
\end{eqnarray}
where $\Gamma$ is the Gamma Function. With this new choice of $\psi$,
$\int_{R_P-4\textrm{R}_{\textrm{H}}}^{R_P+4\textrm{R}_{\textrm{H}}} 2 \pi R \psi(R,R_P,\textrm{R}_{\textrm{H}})~dR \sim
0.96$, so the tail of the function has a negligible contribution in
Eq.~({\ref{eq:mc_acc_int}}) and continue be smooth for a numerical
treatment. We employ a Simpson rule to integrate
Eq.~({\ref{eq:mc_acc_int}}) where at least 10 radial bins between
$R_P-4\textrm{R}_{\textrm{H}}$ and $R_P+4\textrm{R}_{\textrm{H}}$ are considered. Finally, the total planetesimal accretion rate is given by,
\begin{eqnarray}
\frac{d\textrm{M}_{\textrm{C}}}{dt}= \sum_j \frac{d\textrm{M}_{\textrm{C}}^j}{dt}.
\label{eq:mc_acc_tot}
\end{eqnarray}

The rest of the model is the same described in detail in Guilera et al. (2010). 

%______________________________________________________________

\section{Planetesimal fragmentation}

We incorporate a model of planetesimal fragmentation in our model of
giant planet formation which is based on the BOULDER code (Morbidelli et
al., 2009 and supplementary material). This code models the accretion
and fragmentation of a population of planetesimals (as our model 
of giant planet formation starts with an embryo already formed, sourrundig 
by a swarm of planetesimals, i.e. in th oligarchic growth regime, we first 
have into account only the corresponding fragmentation prescriptions, see 
next sections).

According to this model, if $Q_D^*$ is the specific impact energy per
unit target mass (energy required to disperse 50\% of the target mass)
and $Q$ is the collisional energy per unit target mass, the collision
between a target of mass $M_T$ and a projectile of mass $M_P$ (with
$M_P \leq M_T$) gives a remnant of mass $M_R$ which is given by, 
\begin{eqnarray}
M_R = \left\{ 
\begin{array}{cc} 
\left[ -\frac{1}{2} \left( \frac{Q}{Q_D^*} - 1 \right) + \frac{1}{2}
\right]~(M_T + M_P), & \textrm{if}~Q < Q_D^*, \\
 & \\
\left[ -0.35 \left( \frac{Q}{Q_D^*} - 1 \right) + \frac{1}{2}
\right]~(M_T + M_P), & \textrm{if}~Q > Q_D^*. 
\end{array} \right.
\label{eq:MR}
\end{eqnarray}
If $M_R > M_T$ the collision results in accretion. On the other hand, 
if $M_R < 0$ the
target is fully pulverized and its mass is lost. In general, $Q_D^*$ is
function of the radius of the target. However, as the model considers that
the mass of the remnant is function of $(M_T + M_P)$, for consistency,
$Q_D^*$ must be calculated using an effective radius given by, 
\begin{eqnarray}
r_{eff}= \left[ \frac{3(M_T + M_P)}{4 \pi \rho} \right]^{1/3},
\label{eq:radio_efectivo}
\end{eqnarray}
where $\rho$ is the density of the planetesimals. The mass ejected
from the collision --defined as $(M_T + M_P - M_R)$-- is distributed
following a power-law mass distribution $dn/dm \propto m^{-p}$ between the
minimum bin mass considered and the bin mass corresponding to the bigger fragment $M_F$ given by, 
\begin{eqnarray}
M_F= 8 \times 10^{-3} \left[ \frac{Q}{Q_D^*}~e^{-(Q/4Q_D^*)^2} \right]~(M_T + M_P).
\end{eqnarray}
We find that for some super catastrophic collisions (when $M_R \ll M_T+M_P$) occurs that $M_F > M_R$. %(Fig.~\ref{fig:Mt.vs.Q}).  
So, for these collisions we set $M_F= 0.5 M_R$. 

The exponent $p$ of the mass distribution is given by,
\begin{eqnarray}
p= \frac{1}{3}(3-q),
\label{eq:expo_p}
\end{eqnarray}
where $q$ is the exponent of the cumulative power-law distribution, and
is given by,
\begin{eqnarray}
q= -10 + 7 \left( \frac{Q}{Q_D^*} \right)^{0.4}~e^{-(Q/7Q_D^*)}.
\label{eq:exponent-cummulative-powerlaw}
\end{eqnarray}

For the specific impact energy, we adopted the prescription given by Benz \& Asphaug (1999). We used the prescription for basalts at impact velocities of 5~km~$\textrm{s}^{-1}$ given by,
\begin{eqnarray}
Q_D^*=  3.5 \times 10^7 r_p^{-0.38} + 0.3 \rho_p r_p^{1.36}, 
\end{eqnarray}
using a planetesimal density of $\rho_p= 1.5~\textrm{g}~\textrm{cm}^{-3}$. We remark that in this prescription we used the effective radius (given by Eq.~\ref{eq:radio_efectivo}) instead the planetesimal's radius.

In our global model we consider the evolution --by migration,
accretion and fragmentation-- of a population of planetesimals of radii
between 1~cm and $r_p^{max}$ (a free parameter). However, in a
collisional regime the evolution of the population is ultimately
governed by the size distribution of the smallest objects. So, the
truncation in $r_p= 1$~cm can generate the accumulation of spurious
mass in the smaller fragments. To avoid this problem, when we
calculate the fragmentation process, we extrapolate the size distribution
to a minimum fragment size of $r_p= 0.01$~cm. In this way, the mass
ejected from the collision is distributed between the mass bin 
corresponded to $r_p= 0.01$~cm and the bin mass corresponding to
$M_F$ as we mentioned above. This means that we are considering that the
mass distributed below the mass bin corresponding to $r_p= 1$~cm is
lost. Moreover, if the mass bin corresponding to $M_R$ is below the mass
bin corresponding to $r_p= 1$~cm we consider that the target is fully
pulverized.   

The total number of collisions between targets $j$ and projectiles $i$
in a time $\Delta t$ is given by (Morbidelli et al., 2009 and supplementary material),  
\begin{eqnarray}
N_C^{j,i}= P_C^{j,i} n_{p_j} n_{p_i} F_g^{j,i}( r_{p_j} + r_{p_i} )^2
 \Delta~t, 
\end{eqnarray}
where $ P_C^{j,i}$ is the intrinsic collision probability,
$n_{p_j}~(r_{p_j})$ and $n_{p_i}~(r_{p_i})$ are the numbers (radii) of
targets and projectiles, respectively, and $F_g^{j,i}$ is the
gravitational focusing factor. Regarding the time step $\Delta~t$, it
is upper limits by a physical condition. Using that $N_{C_{Tot}}^j$ is the total
number of collisions of the targets $j$ given by, 
\begin{eqnarray}
N_{C_{Tot}}^j = \sum_i N_C^{j,i}= n_{p_j} \sum_i P_C^{j,i} n_{p_i}
 F_g^{j,i}( r_{p_j} + r_{p_i} )^2 \Delta~t,       
\end{eqnarray}  
and defining $\tau_i$ as, 
\begin{eqnarray}
\tau_i= \sum_i P_C^{j,i} n_{p_i} F_g^{j,i}( r_{p_j} + r_{p_i} )^2, 
\end{eqnarray} 
we can put,
\begin{eqnarray}
N_{C_{Tot}}^j = n_{p_j} \tau_i \Delta~t.  
\end{eqnarray}
Then, $N_{C_{Tot}}^j$ can not be greater than $n_{p_j}$, so for our
model we adopt that $\Delta~t < 0.1/\tau_i$. This condition implies that
for our global model the time step can not be greater than
$10^{-4}$~My for $r_p^{max}= 10, 100$~km, and $10^{-5}$~My for
$r_p^{max}= 0.1, 1$~km.

\subsection{Implementation of the fragmentation model}

The evolution of the surface densities of
planetesimals obeys a continuity equation,
\begin{eqnarray}
 \frac{\partial}{\partial t}\left[ \Sigma_p(R,r_p) \right] -
  \frac{1}{R}\frac{\partial}{\partial R}
  \bigg(R\frac{dR}{dt}\Sigma_p(R,r_p)\bigg) = \mathcal{F}(R,r_p), 
\label{eq:evol_dens_planetes}
\end{eqnarray}
wherein $R$ and $r_p$ reference radial and planetesimal size
dependencies, and $\mathcal{F}$ are the sink terms. In ours previous works, we only considered the planetesimal accretion by forming embryos as sink term. With the incorporation of the planetesimal fragmentation, we introduce a new sink
term in Eq.~(\ref{eq:evol_dens_planetes}), which is solved with a full implicit method in finite differences. 

\begin{figure} 
\centering 
\includegraphics[height=0.125\textheight]{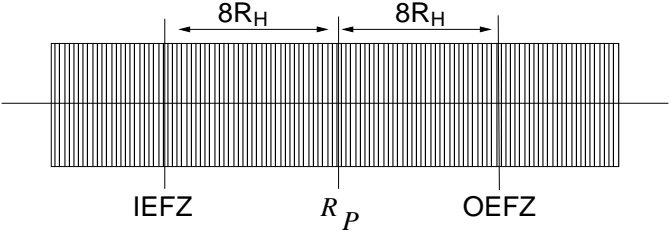} 
\caption{Schematic illustration of the fragmentation zone for an isolated
 planet into the radial grid. IEFZ (OEFZ) represents the inner (outer)
 edge of the fragmentation zone, while $R_P$ is the radial bin
 corresponding to the planet. The fragmentation zone extends 8 Hill
 radii on either side of the planet.}   
\label{fig:zona_frag} 
\end{figure}

To incorporate the fragmentation process into the global model, we define a
fragmentation zone for each embryo wherein we calculate the
collisional process (Fig.~\ref{fig:zona_frag}). This zone extends 8 Hill radii on either
side of the embryo (twice the feeding zone)\footnote{If the
fragmentation zones of two embryo overlap, we only define one
fragmentation zone containing both embryos}. In this way
we reduce the computational cost, and safely guarantee that collisions
that produce fragmentation (craterization or catastrophic collisions)
are within this zone, since they do not extend far away the feeding
zone. In fact, 
the excitations of the eccentricities and inclinations of the
planetesimals (and hence the relative velocities) abruptly decay far away
the feeding zone (Eq.~\ref{eq:evol_ei}). 

In each radial bin of the fragmentation zone, the eccentricities,
inclinations and surface densities for each size of planetesimals are
defined. As we mentioned above, for the calculation of the evolution of
the system mass, we consider a size distribution between 1~cm and a
$r_p^{max}$, where initially the total mass of the system is distributed
in the planetesimals of radii $r_p^{max}$. However, to model the
fragmentation we extrapolate the planetesimal sizes (also the
eccentricities, inclinations and numbers of bodies) to planetesimals of
radius $r_p= 0.01$~cm. In this way, we avoid spurious mass accumulation
in the smaller planetesimals of the distribution.

The implementation methodology is as follows,

\begin{itemize}
\item eccentricities, inclinations and surface densities are defined for 
      each size bin (between $r_p=1$~cm and $r_p= r_p^{max}$) and for
      each radial bin in the fragmentation zone. From the surface densities,
      the number of bodies for each size bin and for
      each radial bin are calculated. With these data, we extrapolate the
      inclinations, eccentricities and number of bodies for the
      corresponding size bins between $r_p = 0.01$~cm and $r_p = 1$~cm,
\item then, we take a target $j$ belonging to the radial bin IEFZ (inner
      edge of the fragmentation zone, Fig.~\ref{fig:zona_frag}), 
\item we take a projectile $i$ (with $r_{p_i} \le r_{p_j}$, i.e. from
      $r_{p_i}= 0.01$~cm to $r_{p_i}= r_{p_j}$) of the
      radial bin IEFZ, and calculate if the orbits of the target and the projectile overlap (taking only into account the eccentricities of the target and projectile), 
\item if the orbits overlap, the number of collisions between targets $j$ and projectiles $i$ are calculated,
\item with this information we can calculate how much mass the
      projectiles $i$ disperse from targets $j$ and how this mass
      (remnant plus fragments) is distributed in smaller planetesimals
      than target $j$,\footnote{Numerically, when a collision occurs, we
      consider that the target $j$ disappears from its corresponding radial bin $a_j$, while the projectile $i$ disappears from its corresponding radial bin $a_i$. On the other hand, the remnant and the fragments are distributed in
      the radial bin $a_j$ corresponding to the target.}
\item this is repeated for all projectiles belonging to the radial bin
      IEFZ. Then we move to the radial bin IEFZ+1, and repeat the
      process for all the projectiles that correspond to target $j$ of the
      radial bin IEFZ. This process is repeated until the radial bin
      OEFZ (outer edge of the fragmentation zone) is achieved, 
\item this process is repeated for all targets $j$ from radial bin IEFZ,
      i.e. from $r_{p_j}=1$~cm to $r_{p_j}= r_p^{max}$, 
\item then, we move to the radial bin IEFZ+1 and repeat the process for all the
      targets. The process is repeated until reaching the radial bin
      OEFZ for the targets. 
\end{itemize}

Completed the process, we have the change in mass (loss and gain) for
each planetesimal size and for each radial bin, product of the
planetesimal fragmentation. With this we can calculate the change in
the surface densities for each planetesimal size inside the fragmentation zone.
 
\section{Results}

\begin{table*}[t]
\caption{Results of the two sets of simulations, considering --and not--
 planetesimal fragmentation (PF). The first column corresponds to the
 disk mass. A value of \# means that we consider a disk \# times more
 massive than the Minimum Mass Solar Nebula (MMSN) of Hayashi
 (1981). $\textrm{M}_{\textrm{C}}$ represents the core mass when the
 planet achieves the critical mass and $\textrm{t}$ represents the time 
 at which it
 occurs. Simulations stopped at $\textrm{t}= 6$~My, so that in this case
 $\textrm{M}_{\textrm{C}}$ represents the core mass at this time. For
 $r_p^{max}= 0.1$~km and for disks 8 and 10 times more massive than the
 MMSN, planetesimal accretion rates become so high that models do not
 converge when planetesimal fragmentation is not considered.}             
\label{tab:general-results}      
\centering     
\small{     
\begin{tabular}{|c|c c|c c|c c|c c|c c|c c|c c|c c|}  % 17 columns 
\hline 
\hline       
\# & \multicolumn{4}{c|}{$r_p^{max}= 0.1$~km} & \multicolumn{4}{c|}{$r_p^{max}= 1$~km} & \multicolumn{4}{c|}{$r_p^{max}= 10$~km} & \multicolumn{4}{c|}{$r_p^{max}= 100$~km} \\  
\hline
 & \multicolumn{2}{c|}{Without PF} & \multicolumn{2}{c|}{With PF} & \multicolumn{2}{c|}{Without PF} & \multicolumn{2}{c|}{With PF} & \multicolumn{2}{c|}{Without PF} & \multicolumn{2}{c|}{With PF} & \multicolumn{2}{c|}{Without PF} & \multicolumn{2}{c|}{With PF} \\
 & $\textrm{M}_{\textrm{C}}$ & $\textrm{t}$ & $\textrm{M}_{\textrm{C}}$ & $\textrm{t}$ & $\textrm{M}_{\textrm{C}}$ & $\textrm{t}$ & $\textrm{M}_{\textrm{C}}$ & $\textrm{t}$ & $\textrm{M}_{\textrm{C}}$ & $\textrm{t}$ & $\textrm{M}_{\textrm{C}}$ & $\textrm{t}$ & $\textrm{M}_{\textrm{C}}$ & $\textrm{t}$ & $\textrm{M}_{\textrm{C}}$ & $\textrm{t}$ \\

 & ($\textrm{M}_{\oplus}$) & ($\textrm{My}$) & ($\textrm{M}_{\oplus}$) & ($\textrm{My}$) & ($\textrm{M}_{\oplus}$) & ($\textrm{My}$) & ($\textrm{M}_{\oplus}$) & ($\textrm{My}$) & ($\textrm{M}_{\oplus}$) & ($\textrm{My}$) & ($\textrm{M}_{\oplus}$ & ($\textrm{My}$) & ($\textrm{M}_{\oplus}$) & ($\textrm{My}$) & ($\textrm{M}_{\oplus}$) & ($\textrm{My}$) \\  
  
\hline                    
 2 & 23.10 & 4.41 & 0.04 & 6.00 & 5.84 & 6.00 & 0.04 & 6.00 & 0.35 & 6.00 & 0.06 & 6.00 & 0.11 & 6.00 & 0.11 & 6.00 \\
\hline 
 4 & 32.23 & 0.39 & 0.09 & 6.00 & 21.37 & 2.99 & 0.08 & 6.00 & 9.17 & 6.00 & 0.18 & 6.00 & 0.56 & 6.00 & 0.52 & 6.00 \\
\hline 
 6 & 35.25 & 0.17 & 0.20 & 6.00 & 27.55 & 0.92 & 0.12 & 6.00 & 21.55 & 3.27 & 0.39 & 6.00 & 2.13 & 6.00 & 1.34 & 6.00 \\
\hline                  
 8 & ($\dots$) & ($\dots$) & 0.45 & 6.00 & 35.73 & 0.33 & 0.15 & 6.00 & 27.77 & 1.64 & 0.67 & 6.00 & 14.99 & 6.00 & 3.58 & 6.00 \\
\hline
 10 & ($\dots$) & ($\dots$) & 0.77 & 6.00 & 45.78 & 0.15 & 0.18 & 6.00 & 32.64 & 0.99 & 1.00 & 6.00 & 25.06 & 4.07 & 7.13 & 6.00 \\
\hline                  
\hline
\end{tabular}
}
\end{table*}

The protoplanetary disk is defined between 0.4~AU and 20~AU, using 2500 radial bins logarithmic equally spaced. We applied our model to study the role of planetesimal fragmentation on giant planet formation. We calculated the in situ formation of an embryo located at 5~AU for different values of the Minimum Mass Solar Nebula (Hayashi, 1981), which is given by
\begin{eqnarray}
\Sigma_p & = & \left\lbrace
\begin{array}{ll}
7.1 \left(\displaystyle{\frac{R}{1~\mathrm{AU}}}\right)^{-3/2}~\mathrm{g}~\mathrm{cm}^{-2}  & \mbox{$R<2.7$~AU}\\
\\
30  \left(\displaystyle{\frac{R}{1~\mathrm{AU}}}\right)^{-3/2}~\mathrm{g}~\mathrm{cm}^{-2}  & \mbox{$R>2.7$~AU},
\end{array}
\right. \\
\Sigma_g & = & 1700 \left(\frac{R}{1~\mathrm{AU}} \right)^{-3/2}~\mathrm{g}~\mathrm{cm}^{-2}, \\
T & = & 280 \left(\frac{R}{1~\mathrm{AU}}\right)^{-1/2}~\mathrm{K}, \\
\rho_g & = & 1.4\times10^{-9} \left(\frac{R}{1~\mathrm{AU}} \right)^{-11/4}~\mathrm{g}~\mathrm{cm}^{-3},
\end{eqnarray}
wherein $\Sigma_p$ and $\Sigma_g$ represent the planetesimal and gaseous surface densities, respectively, $T$ is the temperature profile and $\rho_g$ is the volumetric density of gas at the mid plane of the disk. The discontinuity at $2.7$~AU in the surface density of planetesimals is caused by the condensation of volatiles, the often called snow line. For numerical reasons, and following Thommes et al.~(2003), we spread the snow line with a smooth function, so the surface density of planetesimals is described by
\begin{eqnarray}
\Sigma_p=  & & \left\lbrace 7.1 + \left( 30-7.1\right) \left[\frac{1}{2}\tanh\left(\frac{R-2.7}{0.5}\right) +  \frac{1}{2}\right]\right\rbrace \times \nonumber \\
 & & \left( \frac{R}{1~\mathrm{AU}}\right) ^{-3/2}~\mathrm{g}~\mathrm{cm}^{-2}. 
\end{eqnarray}   

We carried out two different sets of simulations. In the first one, we took into account that the planetesimal surface density evolved only by accretion of planetesimals by the embryo and for the orbital migration of planetesimals. In the other set of simulations the planetesimal surface density evolved by accretion, orbital migration and planetesimal fragmentation.   

Regarding the gaseous component of the disk, Alexander et al. (2006) found that after a few My of viscous evolution the disk could be completely dissipated by photo-evaporation in a time-scale of $10^5$~yr. For simplicity, we considered that the gaseous component of the disk exponentially dissipated in 6~My, when photo-evaporation acts and completely dissipates it. So, we ran our models until the embryo achieved the critical mass (when the mass of the envelope equals the mass of the core) or for 6~My.  

We started our simulations with an embryo of
$0.005~\textrm{M}_{\oplus}$, which has an initial envelope of $\sim
10^{-13}~\textrm{M}_{\oplus}$, immersed in an initial homogeneous
population of planetesimals of radius $r_p^{max}$ (we considered
different values for $r_p^{max}$: 0.1, 1, 10, 100~km). It is important
to remark that our initial conditions correspond to the beginning of the
oligarchic growth. Ormel et al.~(2010), employing statistical
simulations found that, starting with an homogeneous population of
planetesimals of radius $r_0$, the transition from the runaway growth
to the oligarchic growth is characterized by a power-law mass
distribution given by $dn/dm \propto m^{-p} $ (with $p \sim 2.5$),
between $r_0$ and a transition radius $r_{trans}$ for the population of
planetesimals, and isolated bodies (planetary embryos). This implies
that most of the solid mass lies in small planetesimals. For simplicity,
and considering the fact that most of the mass lies in the smaller
planetesimals of the population, we used a single size distribution
instead of a planetesimal size distribution to represent the initial planetesimal population. So, our initial conditions are consistent with the oligarchic growth regime using $r_p^{max}$ as $r_0$.

In this work we want to analyze how planetesimal fragmentation impacts on the process of planetary formation. Accretion collisions between planetesimals are important to study the transition from planetesimal runaway growth to the oligarchic growth. In the oligarchic growth, embryos gravitationally dominate the dynamical evolution of the surrounding planetesimals. As embryos grow they increase the planetesimal relative velocities and collisions between planetesimals result in fragmentation (erosive or disruptive collisions). For these reasons, we focused our analysis in fragmentation collisions. However, as we show in next sections, for some special cases the total planetesimal accretion rates are dominated by the accretion of very small fragments ($r_p \sim 1$~m). For these small fragments, collisions between them (and obviously with smaller fragments) result in accretion. So, for these special cases we also calculated coagulation between planetesimals (see next sections for the detail discussion of this topic). 

As we mentioned in previous section, when planetesimal fragmentation is considered (when not, we used a single size distribution to represent the planetesimal  population) we used a discrete size grid between 1~cm and $r_p^{max}$ to represent the continuous planetesimal size distribution where the fragments, products of planetesimal fragmentation, are distributed. The size step is given by Eq.~(\ref{eq:size-step}). As we used the same size step independently of the value of $r_p^{max}$, this implies that we used 21 size bins for $r_p^{max}= 0.1$~km, 26 size bins for $r_p^{max}= 1$~km, 31 size bins for $r_p^{max}= 10$~km, and 36 size bins for $r_p^{max}= 100$~km. In all cases, the initial total mass of solids along the disk remains in the size bins corresponding to $r_p^{max}$. The collisional evolution of planetesimals is the mechanism that regulates the exchange of mass between the different size bins.  

In previous works (Guilera et al. 2010, 2011) we showed that the in situ
simultaneous formation of solar system giant planets occurred in a
time-scale compatible with observed estimations only if most of the
solid mass accreted by the planets remains in small planetesimals ($r_p <
1$~km). Recently, Fortier et al. (2013) studied in detail the role of planetesimal sizes in planetary population synthesis. They found that including oligarchic growth the formation of giant planets using big planetesimals ($r_p \sim 100$~km) is unlikely, and only if most of the mass of the system remains in small objects ($r_p \sim 0.1$~km) cores grow enough to form giant planets. However, small planetesimals have smaller specific impact energies per unit target mass. So, small planetesimals suffer catastrophic collisions for smaller collisional energies.

\begin{figure}
  \centering
  \includegraphics[angle=270, width= 0.49\textwidth]{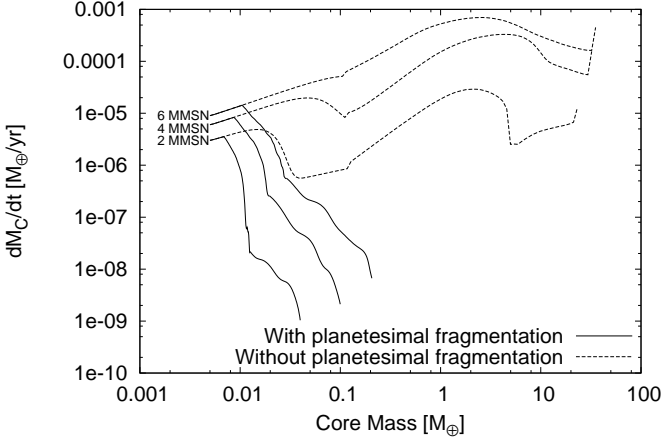}
  \caption{Total planetesimal accretion rates, for three different disks, as function of the core mass. The case where planetesimal fragmentation is not considered is shown in dotted lines. For this case, the critical mass is achieved for the three different disks in less than 6~My. When planetesimal fragmentation is incorporated the total planetesimal accretion rates significantly drop. Despite the lower values of the core masses, the fragmentation process truncated the embryo growth.}
  \label{fig:tasas-rp-0.1km}
\end{figure}

The results of the two sets of simulations are summarized in
Tab.~\ref{tab:general-results}. For smaller planetesimals ($r_p = 0.1$~km), the collisional evolution completely inhibited planetary formation. When planetesimal fragmentation is not taken into account the embryo is able to achieve the critical mass for 2 to 10 values of the MMSN before the dissipation of the nebula\footnote{For 8 and 10 values of the MMSN
planetesimal accretion rates become so high that models do not
converge.}. Moreover, for disks more massive than 4 times the MMSN the
times at which embryo achieves the critical mass are very short
($\lesssim 0.5$~My), while the critical masses are high ($\gtrsim
30~\textrm{M}_{\oplus}$). When we included planetesimal fragmentation
the picture drastically changed. For none of the analyzed cases, the
critical mass is reached. Moreover, even for the case of the more
massive disk (10 MMSN), the planetary embryo is not able to achieve one
Earth mass. In Fig.~\ref{fig:tasas-rp-0.1km}, we show the total
accretion rate of planetesimals as function of the core mass for the case of $r_p^{max}=0.1$~km. When planetesimal fragmentation is incorporated the total planetesimal accretion rate is completely inhibited despite the small values of the core masses. The small differences between the disks at which values of the core mass planetesimal fragmentation starts to decrease the accretion rates are because of the amount of gas of the different disks. The more massive disk, the more damping in the planetesimal eccentricities and inclinations, so planetesimal relative velocities are lower for the same value of the core mass.  

\begin{figure}
  \centering
  \includegraphics[angle=270, width= 0.475\textwidth]{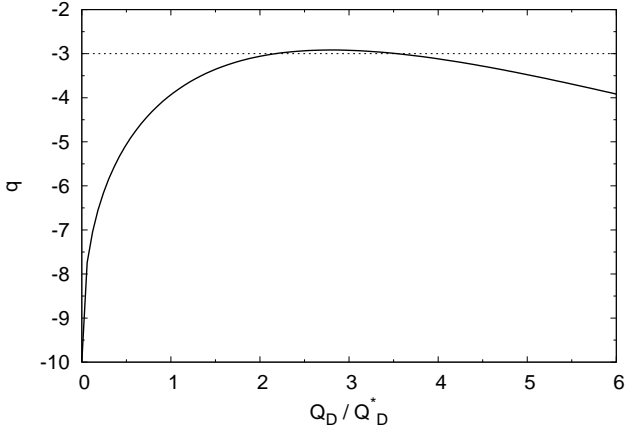}
  \caption{Exponent of the cumulative power law distribution as function of $Q / Q_D^*$ given by the fragmentation model. The value $q= -3$ implies that $p= 2$ (Eq.~\ref{eq:expo_p}). This means that the mass lost in a collision is distributed homogeneously between the fragments.} 
  \label{fig:expo_q}
\end{figure}

These drops in the planetesimal accretion rates are due to
a drastic diminution in the mean value of the surface density of
planetesimals of radius 0.1~km in the feeding zone, respect to the case
where planetesimal fragmentation is not taken into account.  
\begin{figure}
 \centering
 \includegraphics[angle=270, width= 0.49\textwidth]{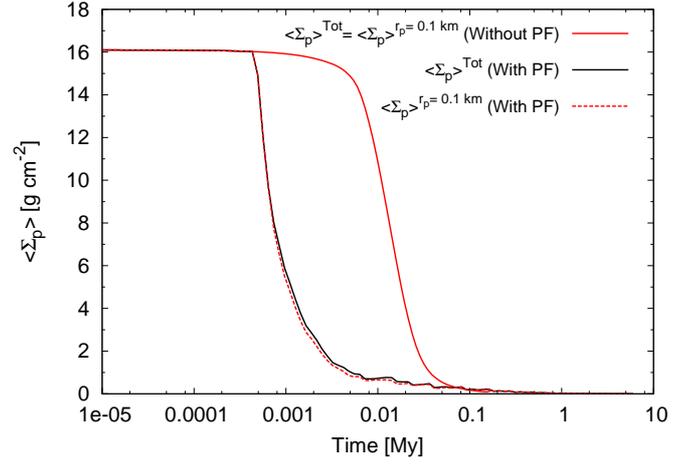}
 \caption{Time evolution of the mean value of the surface density of
 planetesimals in the feeding zone for a disk 6 times more massive than
 the MMSN. For the case wherein planetesimal fragmentation is not
 considered (red solid curve), the total surface density is the corresponding to
 planetesimals of radius 0.1~km. This is not the case when planetesimal
 fragmentation is considered (solid black curve). However, the total
 surface density is dominated by the surface density of planetesimals of
 0.1~km (dashed red curve). Color figure only available in the electronic version.} 
 \label{fig:densi-rp-0.1km}
\end{figure}   
In Fig.~\ref{fig:densi-rp-0.1km}, we plot the time evolution of the mean
value of the surface density of planetesimals in the feeding zone for a
disk 6 times more massive than the MMSN. When we included planetesimal
fragmentation the total planetesimal surface density significantly drops
(solid black curve). It is important to remark that for the case where
planetesimal fragmentation is not taken into account (solid red curve),
the total planetesimal surface density corresponds to the surface
density of planetesimals of radius 0.1~km. This is not the case when we
considered planetesimal fragmentation. However, as we can see in
Fig.~\ref{fig:densi-rp-0.1km} the surface density of planetesimals of
radius 0.1~km (dashed red curve) is almost the total planetesimal
surface density. This is because of two effects. First, most of the mass
distributed in the fragments, products of the collisions, is lost below
the size bin corresponding to 1~cm. As we mentioned above, we considered
that the mass loss in collisions is distributed following a
power law mass distribution between fragments of $r_p=0.01$~cm and the
biggest fragment of mass $M_F$. So, if the exponent $p$ of the mass
distribution ($dn/dm \propto m^{-p}$) is greater than 2 most of the mass
is distributed in the smaller fragments, and if $p < 2$ most of the mass
is distributed in bigger fragments. When $p > 2$ most of the mass is
distributed in fragments lower than 1~cm, so this mass is lost in our
model. In Fig.~\ref{fig:expo_q}, we show the values of $q$, the exponent
of the cumulative power-law distribution
(Eq.~\ref{eq:exponent-cummulative-powerlaw}), as function of $Q /
Q_D^*$. We can see that except for values between $2 \lesssim Q / Q_D^*
\lesssim 3.5$, $q$ is always lower than -3. This mean that in general
$q<-3$, so $p>2$ (see Eq.~\ref{eq:expo_p}) and most of the mass
distributed in fragments is lost. It is important to remark that
for such step distributions, the integration over the mass, between $m=
0$ and $m= M_F$, diverges. In the Boulder code, this problem is solved
using that these mass distributions are valid between $m= m_t$ ($m_t$
represents a transition mass) and $m=M_F$ and using an ad-hoc mass
distribution with an exponent $p= 11/6$ between  $m=0$ and $m=m_t$. In
this approach, most of the mass loss in collisions is distributed in the
radial bin correspondig to $r_t$ (the transition radius corresponding to
$m_t$). However, we don't follow this approach. We truncate the mass
distribution in a minimum mass, the one corresponding to $r_p= 0.01$~cm
(we tested lower values than $r_p= 0.01$~cm founding analogous
results). This is an alternative approach, adopting a fixed value of $r_t= 0.01$~cm and calculating the exponent $p$, such that the integration over the
mass between $m=0$ and $m= M_F$, converges.\footnote{For some
simulations, we tested the approach given by the Boulder code founding
that $r_t$ is always lower than 1~cm. However, we note that we had to
adopt that there is only one body of mass $M_F$ to calculate all the
free parameters in the resulting integral equation of the mass.} In this way, we always guarantee --for each collision-- that the mass distributed between $r_p= 0.01$~cm and the radial bin corresponding to planetesimals of mass $M_F$ is never greater than the mass ejected from the collision ($M_T + M_P - M_R$).  

\begin{figure}
 \centering
 \includegraphics[angle=270, width= 0.475\textwidth]{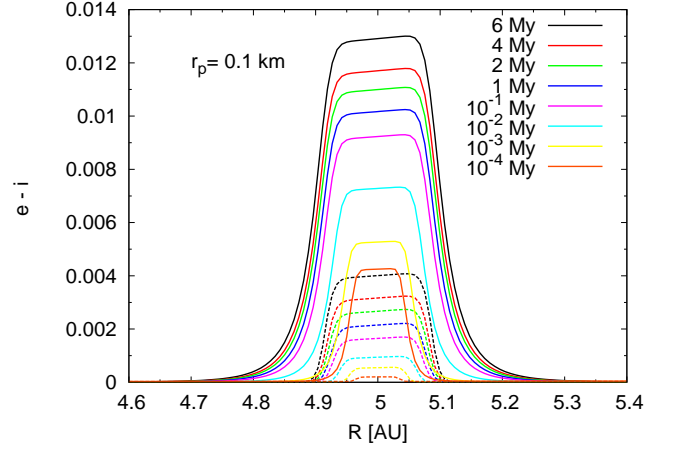}
 \caption{Time evolution of the radial profiles of the eccentricities (solid lines) and inclinations (dashed lines) at the embryo's neighborhood for planetesimals of 0.1~km of radius. As time advance, the planet's gravitational excitation increases the eccentricities and inclinations of the planetesimals. It is also clear that planetesimals do not reach equilibrium values ($\beta= i/e \ne 0.5$). Color figure only available in the electronic version.} 
 \label{fig:exce_incli-rp0.1km}
\end{figure}

\begin{figure}
 \centering
 \includegraphics[angle=270, width= 0.475\textwidth]{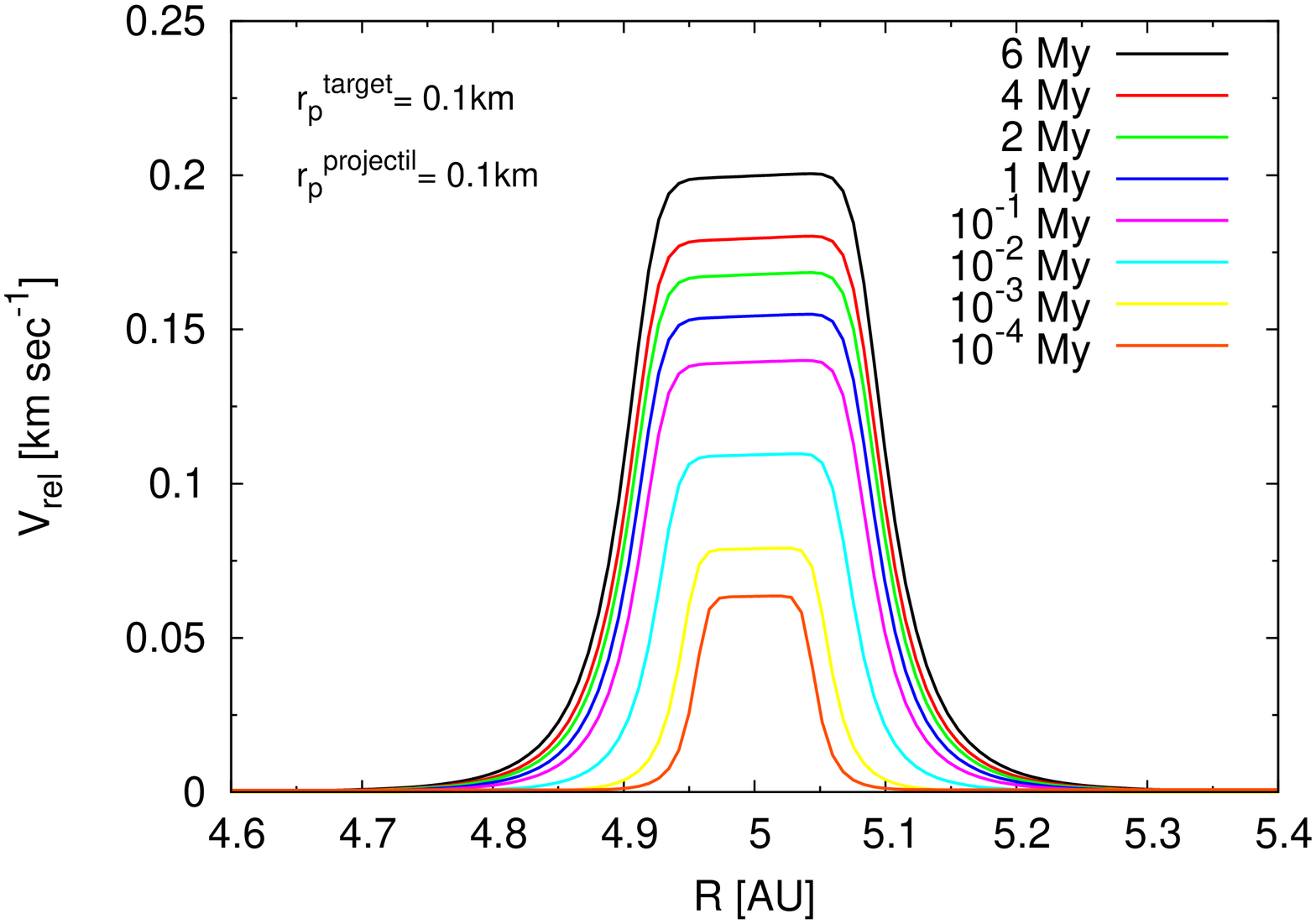}
 \includegraphics[angle=270, width= 0.475\textwidth]{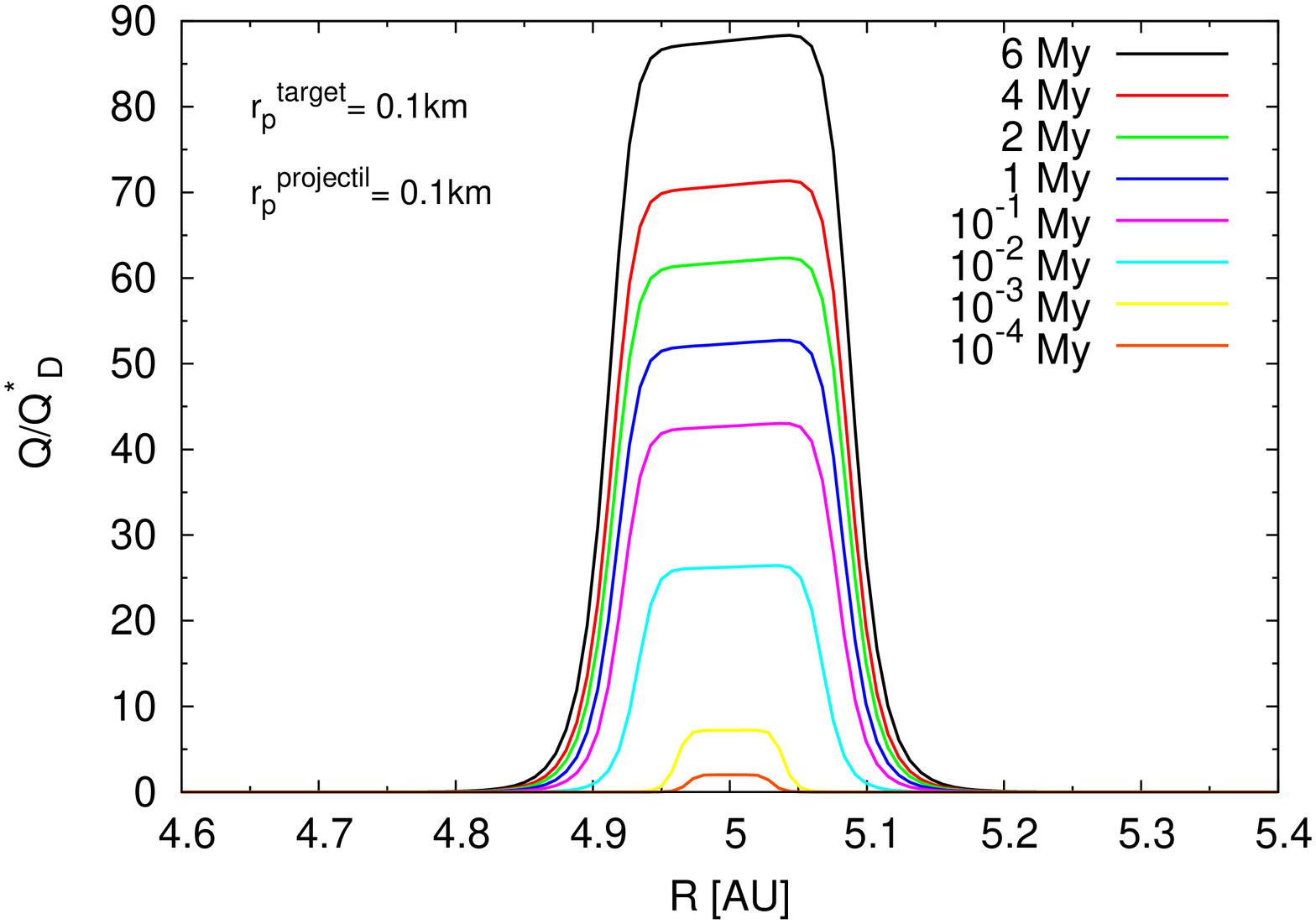} 
 \caption{Time evolution of the radial profiles for the relative
 velocities (top) and for the ratio $Q/Q_D^*$ (bottom) for targets and projectiles of radius 0.1~km when both belong to the same radial bin. Instead the relative velocities are always lower than $1~\textrm{km}~\textrm{sec}^{-1}$, the ratio $Q/Q_D^*$ is much greater than unity at very early times. So, collisions become quickly highly catastrophic at early times. Color figure only available in the electronic version.}  
 \label{fig:Vrel-Q-0.1km}
\end{figure}

The second effect is that the remnants, products of the collisions
between planetesimals of radius 0.1~km, are quickly pulverized. As
these planetesimals initially contain all the mass of the system, the
pulverization of the remnants implies a high loss of mass. For example, in Fig.~\ref{fig:exce_incli-rp0.1km} we can see the radial profiles of the 
eccentricities and inclinations at different times at the embryo's
neighborhood. The gravitational perturbations of the planet increase the eccentricities and inclinations of the planetesimals near the planet's location. From this profile we can analyze the relative velocities, and the ratio $Q/Q_D^*$, when targets and projectiles belong to the same radial bin. Fig.~\ref{fig:Vrel-Q-0.1km} shows the time evolution of the radial profiles for the relative velocities (top panel) and for the ratio $Q/Q_D^*$ (bottom panel) when targets and projectiles belong to the same radial bin. From Eq.~(\ref{eq:MR}), if the ratio $Q/Q_D^*$ corresponding to a collision is greater than $\sim 2.5$, the remnant of such collision is pulverized. As we can see from
Fig.~\ref{fig:Vrel-Q-0.1km}, the collisions between planetesimals of 0.1~km
of radius become quickly supercatastrophics and the mass of the remnants, 
products of them, is lost.  

\begin{figure}
  \centering
  \includegraphics[angle=270, width= 0.475\textwidth]{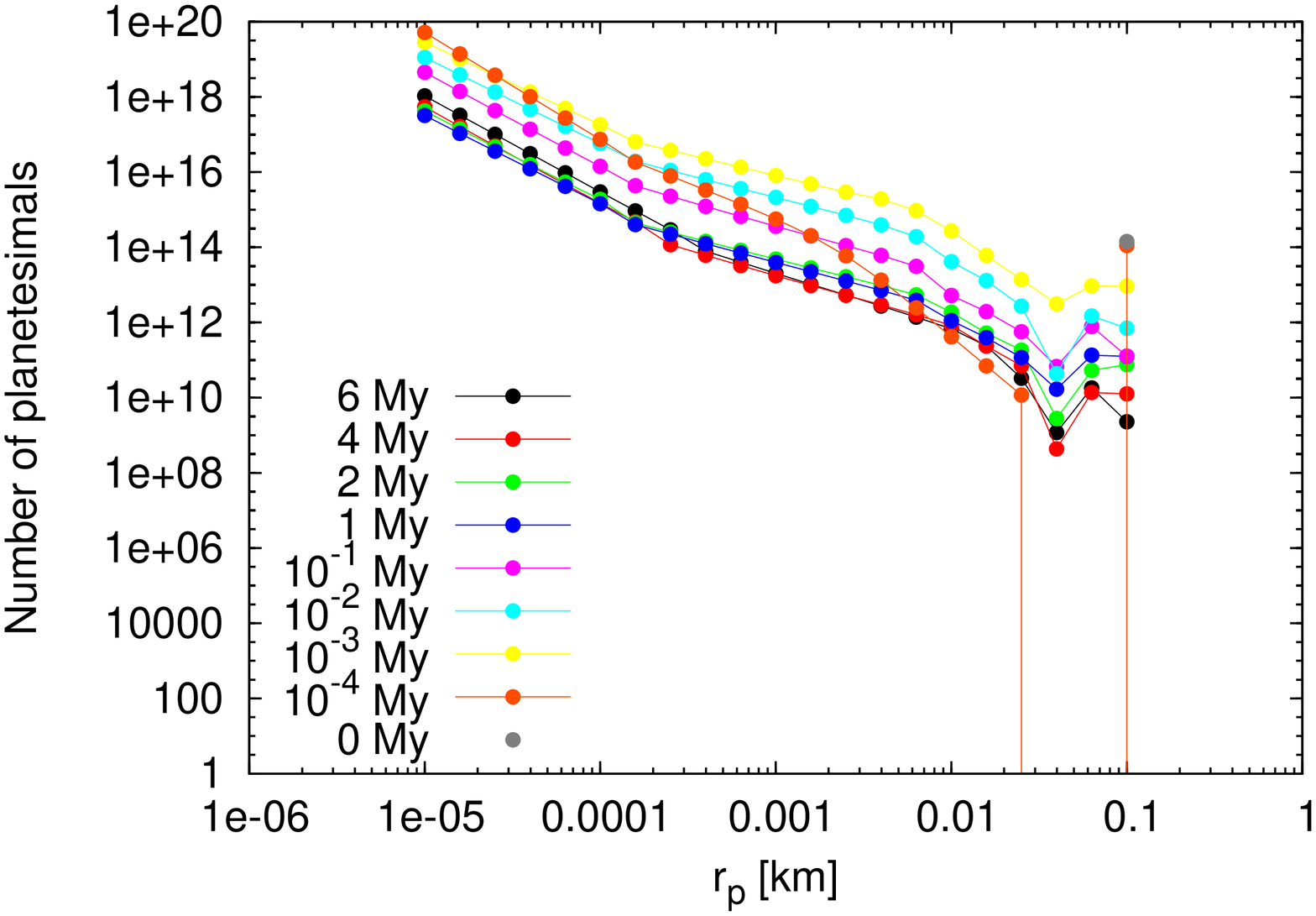}
  \includegraphics[angle=270, width= 0.475\textwidth]{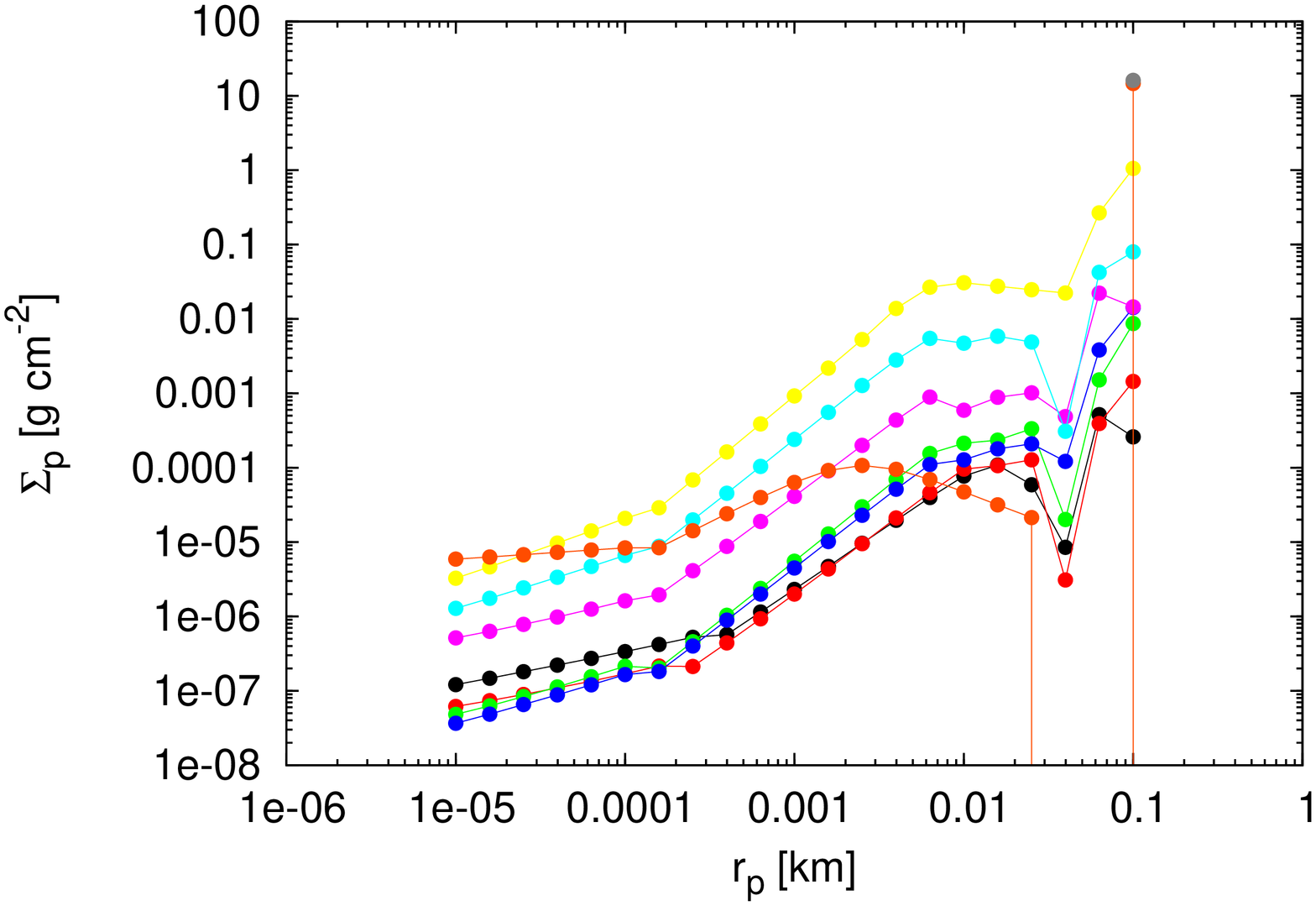}
  \caption{Time evolution of the number of planetesimals (top) and
 planetesimal surface density (bottom) as function of planetesimal
 radius at the embryo's radial bin. The number of planetesimals evolves
 by accretion by the embryo, migration and planetesimal fragmentation. The generation of fragments (produced by the collisional evolution) does not compensate the large decrease of planetesimals of radius 0.1~km. Color figure only available in the electronic version.}
  \label{fig:numero_cuerpos_0.1km}
\end{figure}

Finally, in Fig.~\ref{fig:numero_cuerpos_0.1km} we show the time
evolution of the number of planetesimals and the planetesimal surface
densities at the embryo's radial bin. We can see that the number
of planetesimals of radius 0.1~km is quickly reduced by the collisional
evolution. We also can see that the generation of fragments does not
compensate the diminish of planetesimals of radius 0.1~km. In fact, the values
of the planetesimal surface densities for planetesimals lower than
0.1~km are always $\ll 1~\textrm{gr}~\textrm{cm}^{-2}$.  

We found similar results for the others values of $r_p^{max}$
considered. Collisions between planetesimals of radius $r_p^{max}$
become supercatastrophic and significantly reduce the total
planetesimal accretion rates. This effect, combined with the fact that
most of fragment mass is deposited in size bins lower to the
one corresponding to $r_p= 1$~cm, caused that the formation of a core
able to reach the critical mass is inhibited. 

\begin{figure}
  \centering
  \includegraphics[angle=270, width= 0.475\textwidth]{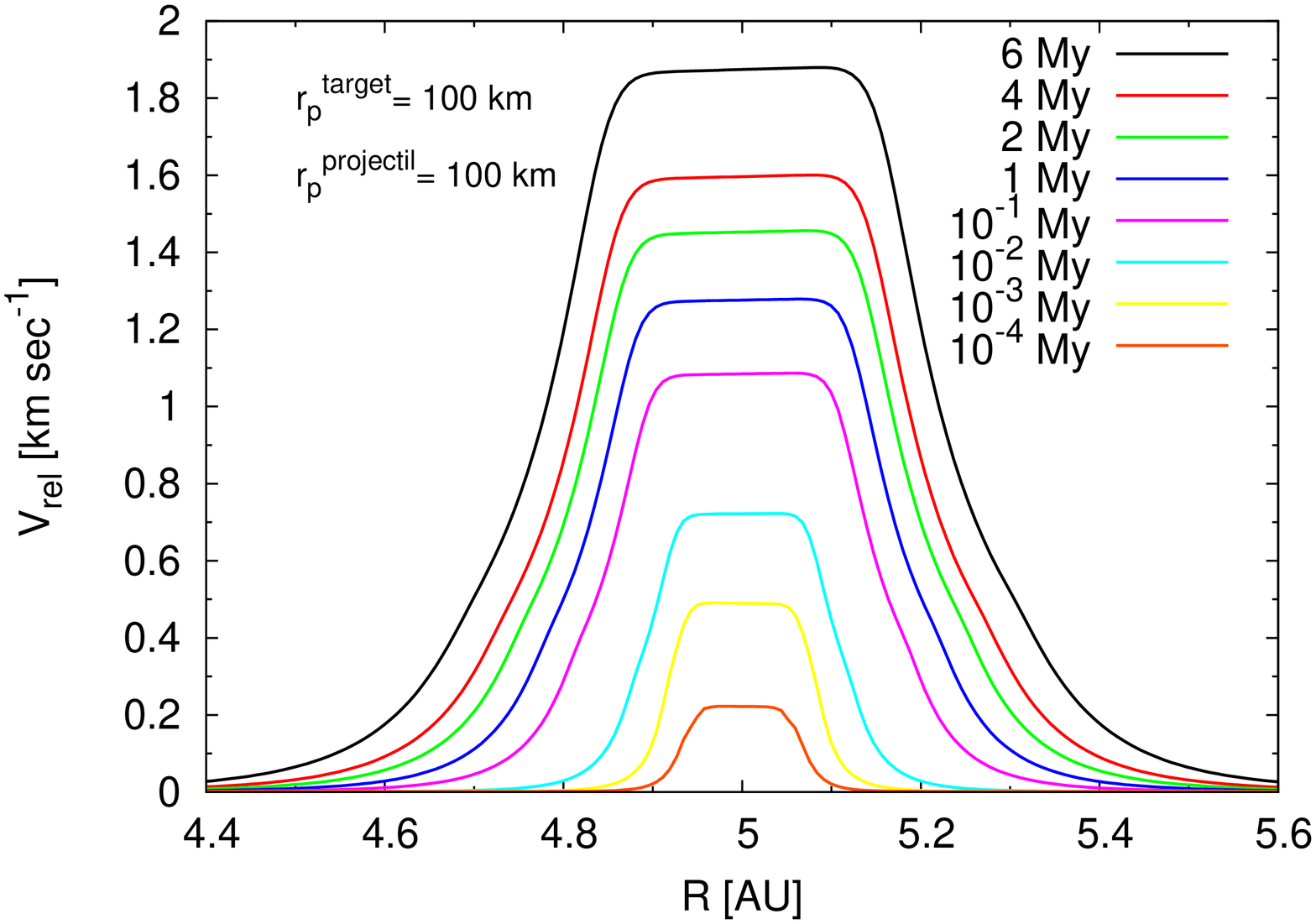}
  \includegraphics[angle=270, width= 0.475\textwidth]{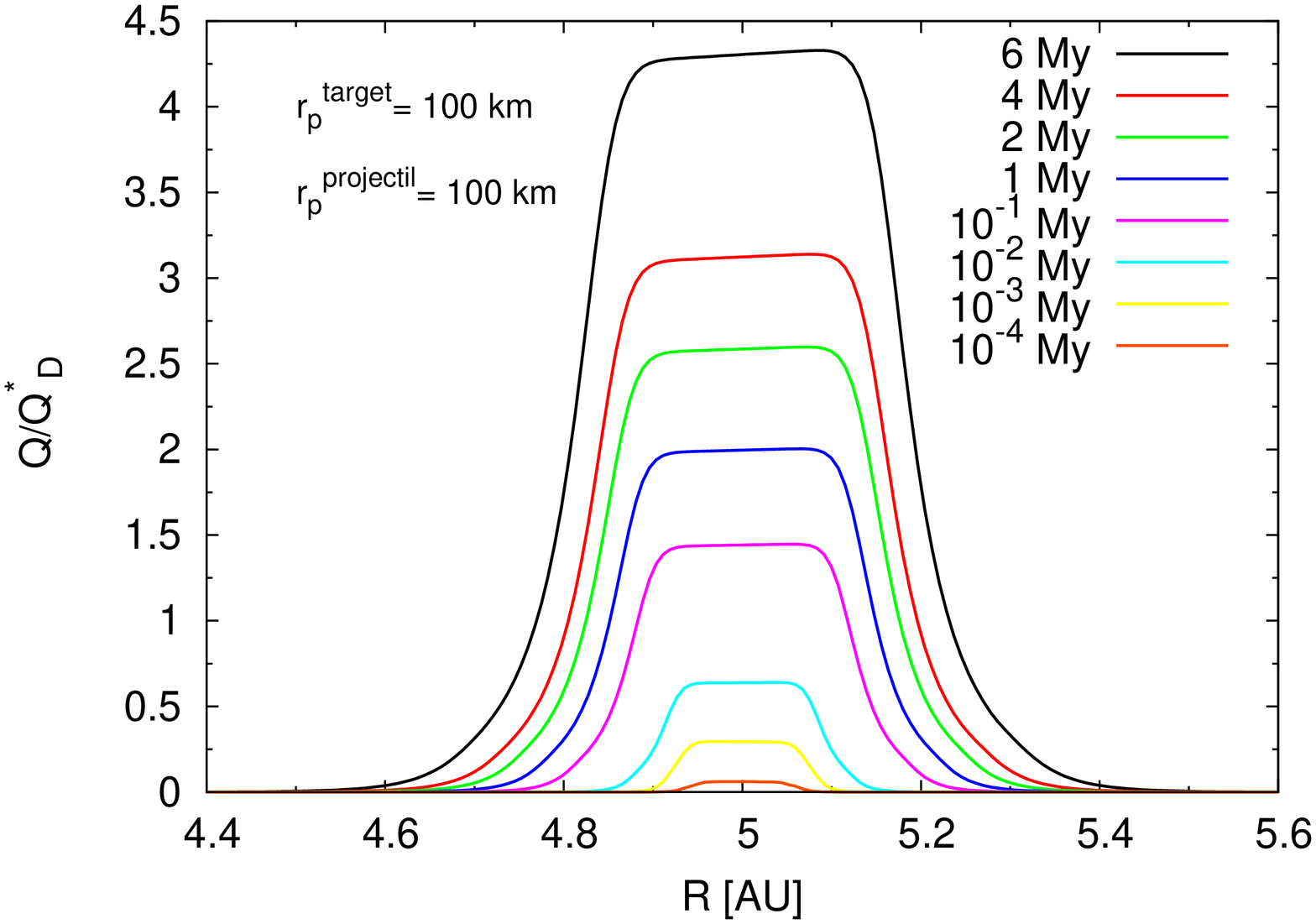}
  \caption{Time evolution of the relative velocity profiles (top) and the ratio $Q / Q_D^*$ profiles (bottom) for targets and projectiles of radii 100~km. The profiles represent the case of $r_p^{max}= 100$~km and a disk 6 times more massive than the MMSN. Despite the larger values of the relative velocities of planetesimals of $r_p= 100$~km, the ratio $Q / Q_D^*$ is lower than small planetesimals. This is because the specific impact energy ($Q_D^*$) for planetesimals of $r_p= 100$~km is approximately three order of magnitude greater than the corresponding to planetesimals of $r_p= 0.1$~km. Color figure only available in the electronic version.}
  \label{fig:Vrel-Q-100km}
\end{figure}

\begin{figure}
 \centering
 \includegraphics[angle=270, width= 0.475\textwidth]{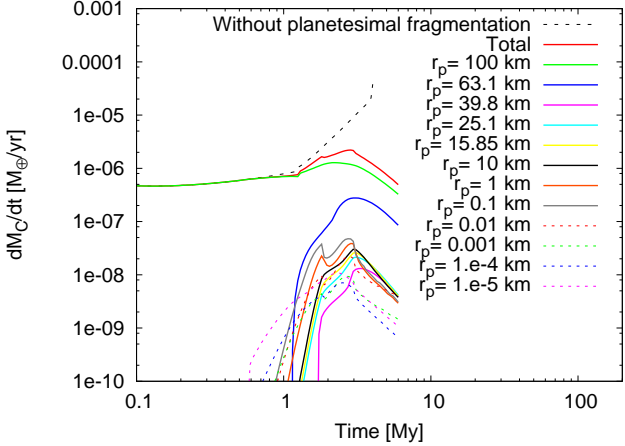}
 \caption{Planetesimal accretion rates as function of time for an embryo located at 5~AU in a disk 10 times more massive than the MMSN, and where initially all the solid mass is deposited in planetesimals of 100~km of radius. Black dashed line corresponds to the case wherein planetesimal fragmentation is not considered. Red solid line corresponds to the total planetesimal accretion rate when planetesimal fragmentation is considered. The accretion rate of planetesimals of radius 100~km is plot in green solid line. We also plot the accretion rates for other planetesimale sizes. Color figure only available in the electronic version.}
 \label{fig:100km-10NM-CB-accretion-rates}
\end{figure}

Only for $r_p^{max}= 100$~km and massive disks, we could form cores with masses greater than one Earth mass. This is because instead that relative velocities for bigger planetesimals are higher, the ratio $Q / Q_D^*$ is lower than small planetesimals (Fig.~\ref{fig:Vrel-Q-100km}). In Fig.~\ref{fig:100km-10NM-CB-accretion-rates}, we plot the planetesimal accretion rates for case of $r_p^{max}= 100$~km and a disk 10 times more massive than the MMSN. In red solid line, we plot the total planetesimal accretion rate. We can see how this accretion rate abruptly drops in comparison with the case wherein planetesimal fragmentation is not considered (black dashed line) because of the drop of the accretion of planetesimals of radius 100~km. We can see too that the accretion of fragments does not compensate the drop in the accretion of planetesimals of radius 100~km. 

\begin{figure}
 \centering
 \includegraphics[angle=270, width= 0.475\textwidth]{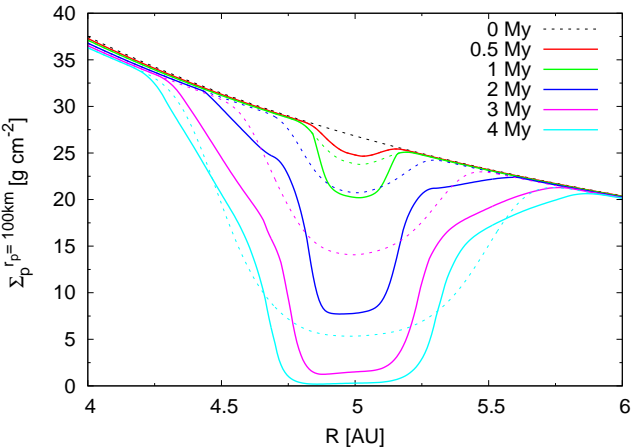}
 \caption{Time evolution of the radial profiles of the surface density of planetesimals of 100~km of radius for a disk 10 times more massive than the MMSN. Solids lines represent the case wherein planetesimal fragmentation is considered, while dashed lines represent the case in which planetesimal fragmentation in not taken into account. For this last case, the planet achieve the critical mass at 4.07~My. When planetesimal fragmentation is considered, the planet did not achieve the critical mass. Color figure only available in the electronic version.}
 \label{fig:densi-comparacion-100km-10NM}
\end{figure}

Finally, in Fig.~\ref{fig:densi-comparacion-100km-10NM} we show the time evolution of the radial profiles of the surface density of planetesimals of 100~km of radius, for the case wherein planetesimal fragmentation is considered (solid lines) and when planetesimal fragmentation is not considered (dashed lines). We can see that the profiles are the same at 0.5~My. At 1~My we can see an evident diminution in the planetesimal surface density around the planet's location (5~AU) for the case wherein planetesimal fragmentation is considered. This diminution in the surface density of planetesimals of 100~km of radius is due to planetesimal fragmentation, but not by planetesimal accretion by the embryo. We can see from Fig.~\ref{fig:100km-10NM-CB-accretion-rates}, that at this time the accretion rate of planetesimals of 100~km of radius is practically the same that the corresponding to the case wherein planetesimal fragmentation is not considered. As time advance, the diminution in the planetesimal surface density around the planet's location becomes greater, for the case wherein planetesimal fragmentation is considered. Finally, at 4 My, the planetesimal surface density is almost zero around the planet's location. We also can see that the loss of mass by planetesimal fragmentation is greater near the planet's location and diminishes far away the location of the planet. 

We want to remark that we found that our results are insensitives for
the numbers of radial and size bins. We tested our results with the
double of radial and size bins obtaining analogues results.  

\subsection{About the distribution of fragments}

As we mentioned in previous sections, most of the fragment mass
produced by the collisions between planetesimals is deposited in size
bins lower than the corresponding to 1~cm, because the exponent $p$ of the
power-law of the fragment mass distribution is generally greater than
2. However, other works suggested that the exponent $p$ should be in a range between 1 and 2. This implies that most of the fragment mass is deposited in bigger fragments and the loss of mass is much lower than in our model. Kobayashi \&
Tanaka (2010) developed a similar fragmentation model (in a qualitative
way about the outcome of a collision) where the mass of the fragments is also distributed following a power-law distribution ($dn/dm \propto m^{-p}$). In Kobayashi et al. (2010; 2011) and Ormel \& Kobayashi (2012), this model is applied to
study planetary formation adopting a value of $p= 5/3$. These works
found that in general planetesimal fragmentation inhibits the formation
of massive cores. Only for massive disks, it is possible the formation of
cores with masses greater than $10~\textrm{M}_{\oplus}$ and only if big
planetesimals are considered ($r_p \geq 100$~km).   

We apply our model using a fixed exponent $p= 5/3$ for the fragments power-law mass distribution. We run again the simulation for a disk 10
times more massive than the MMSN using an initial population of
planetesimals of radius 100~km (our best case). Using a fixed exponent
$p= 5/3$ we found that planetesimal fragmentation favors the formation
of a massive core. For this simulation, we found that the embryo
achieved the critical mass at 3.61~My ($\sim 0.5$~My less than the case
without planetesimal fragmentation) with a core of
$18.58~\textrm{M}_{\oplus}$ ($\sim 6.5~\textrm{M}_{\oplus}$ lower than
the case without planetesimal fragmentation). 

\begin{figure}
  \centering
  \includegraphics[angle=270, width=
 0.475\textwidth]{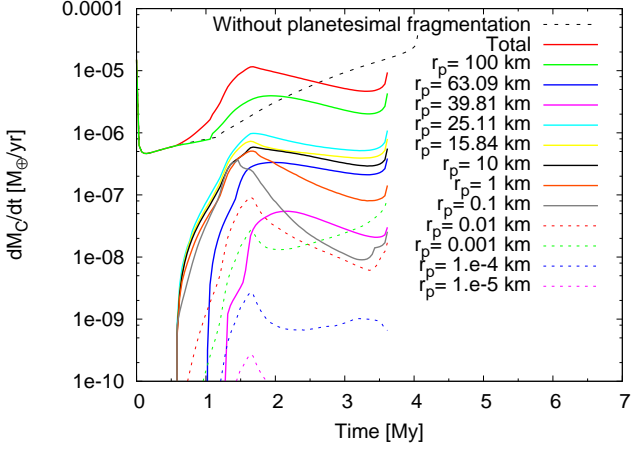}
  \caption{Planetesimal accretion rates as function of
 time for an embryo located at 5~AU in a disk 10 times more massive than
 the MMSN, and where initially all the solid mass is deposited in
 planetesimals of 100~km of radius. Dashed black curve represents the
 total planetesimal accretion rate for the case where planetesimal
 fragmentation is not considered. Red solid curve represents the total
 planetesimal accretion rate for the case in which $p= 5/3$ is a fixed
 value for all collisions. We also plot the planetesimal accretion rates for the
 different planetesimal sizes for the case wherein $p= 5/3$. Color
 figure only available in the electronic version.} 
  \label{fig:tasas-rp100km}
\end{figure}

\begin{figure}
  \centering
  \includegraphics[angle=270, width= 0.475\textwidth]{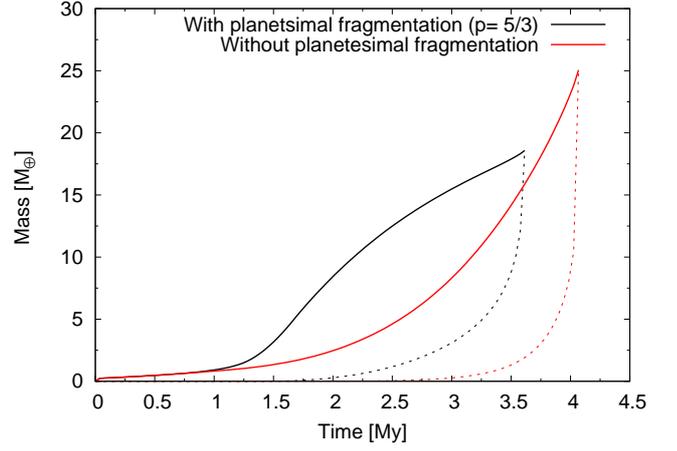}
  \includegraphics[angle=270, width= 0.475\textwidth]{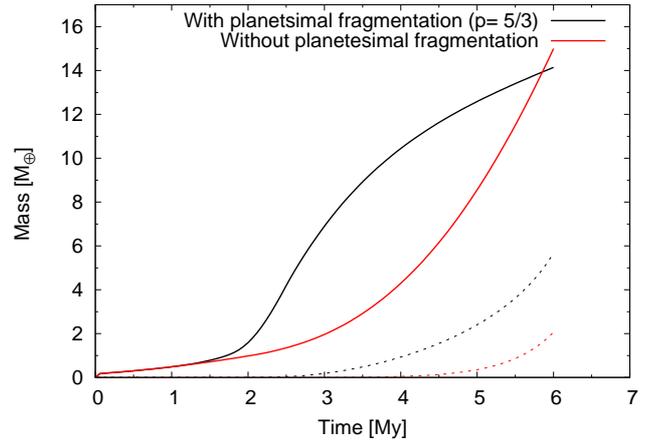}
  \caption{Core masses (solid lines) and envelope masses (dashed lines) as
 function of time for the cases with planetesimal fragmentation
 considering $p= 5/3$ (black lines) and without planetesimal fragmentation (red
 lines), for a disk 10 times more massive than the MMSN (top) and a disk 8
 times more massive than the MMSN (bottom). For this last case, the planets didn't achieve the critical mass after 6~My of evolution. Color figure only available in the electronic version.} 
\label{fig:masas-rp100km}
\end{figure}

In Fig.~\ref{fig:tasas-rp100km} we plot the time evolution of the total
planetesimal accretion rates for the case without planetesimal
fragmentation and the case of $p= 5/3$, and the planetesimal accretion rates for different planetesimal sizes for the case of $p= 5/3$. For the case in which
planetesimal fragmentation is not considered, the planetesimal
accretion rate of planetesimals of radius 100~km corresponds to the total
planetesimal accretion rate. However, this is not the case when
planetesimal fragmentation is considered. Between $\sim 0.5$~My and
$\sim 1$~My the accretion rate of planetesimals of $r_p=100$~km is
slightly lower than the corresponding to the case without
fragmentation. However, the total planetesimal accretion rate is
greater. This is because the accretion of fragments, especially for the
accretion of fragments between $\sim 0.1$~km and $\sim 25$~km. Then, the accretion rate of planetesimals of $r_p=100$~km is
increased at $\sim 1$~My. This is because at this time the ratio
$\tilde{\textrm{R}}_{\textrm{C}} / \textrm{R}_{\textrm{C}}$ becomes greater than unity, so the enhanced in the capture cross section due to the embryo's envelope for planetesimals of radius
100~km make more efficient the accretion of such planetesimals. The
total planetesimal accretion rate remains greater than the corresponding
to the case without fragmentation until $\sim 2.5$~My. This excess in the total planetesimal accretion rate
produces that at $\sim 2.5$~My the planet has a core of $\sim
12.5~\textrm{M}_{\oplus}$ (Fig.~\ref{fig:masas-rp100km} top). At the
same time, the planet corresponding to the case in which planetesimal
fragmentation is not considered has a core of $\sim
4.5~\textrm{M}_{\oplus}$. After 2.5~My, the total planetesimal
accretion rate decreases because of planetesimal fragmentation until the
planet reaches the gaseous runaway phase.

\begin{figure}
  \centering
  \includegraphics[angle=270, width= 0.475\textwidth]{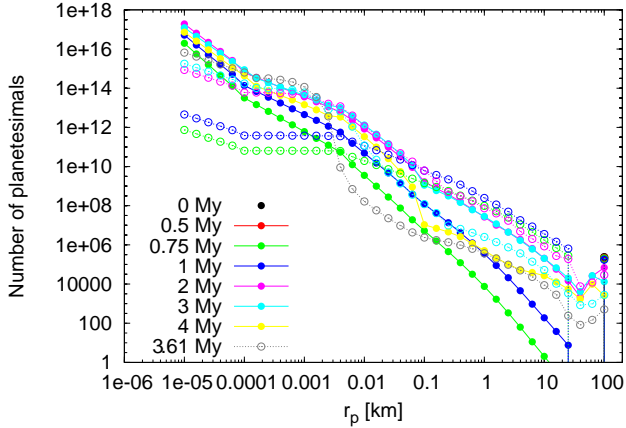}
 \caption{Time evolution of the number of planetesimals adopting
 $r_p^{max}= 100$~km and a disk 10 times more massive than the
 MMSN. Filled circles represent the case wherein the mass loss in
 collisions is deposited in smaller fragments, while open circles
 represent the case wherein the mass loss in collisions is deposited in
 bigger fragments. For this last case, the planet achieved the critical mass at $\sim 3.61$~My. Color figure only available in the electronic version.}
\label{fig:numero-cuerpos-100km-comparacion}
\end{figure}

In Fig.~\ref{fig:numero-cuerpos-100km-comparacion}, we show a comparison of the time evolution of the number of planetesimals at the planet's radial bin, for the cases wherein the exponent of the power-law that represents the planetesimal mass distribution has the constant value of $5/3$ (open circles) and wherein such exponent is calculated via Eq.~(\ref{eq:expo_p}) (filled circles). We can see how the number of fragments between $\sim 0.1$~km and $\sim 25$~km is much greater in the first case for 0.75~My, 1~My, and 2~My. This is because for this case, the mass loss in collisions is distributed in bigger fragments. The accretion of this fragments favors the formation of a massive core. Then, for 3~My, the number of fragments between $\sim 0.1$~km and $\sim 25$~km is lower than the case wherein the mass loss in collisions is distributed in smaller fragments. But this is because of the accretion of such fragments. Finally, for the first case the formation of the core occurs at $\sim 3.61$~My.

Similar behavior occurs for less massive disks. However, for a disk 8
times more massive than the MMSN the planet did not reach the critical
mass. After 6 My of evolution, the planet achieved a total mass of $\sim
20~\textrm{M}_{\oplus}$ ($\sim 14~\textrm{M}_{\oplus}$ for the core and
$\sim 6~\textrm{M}_{\oplus}$ for the envelope, Fig.~\ref{fig:masas-rp100km} bottom ). Although this final core is lower than the corresponding to the case without planetesimal fragmentation
(Tab.~\ref{tab:general-results}), at 4~My the planet has a core slightly
greater than $10~\textrm{M}_{\oplus}$ (in the case without planetesimal fragmentation the planet has a core of $\sim 4~\textrm{M}_{\oplus}$ at the same time).  

\begin{figure}
  \centering
  \includegraphics[angle=270, width= 0.45\textwidth]{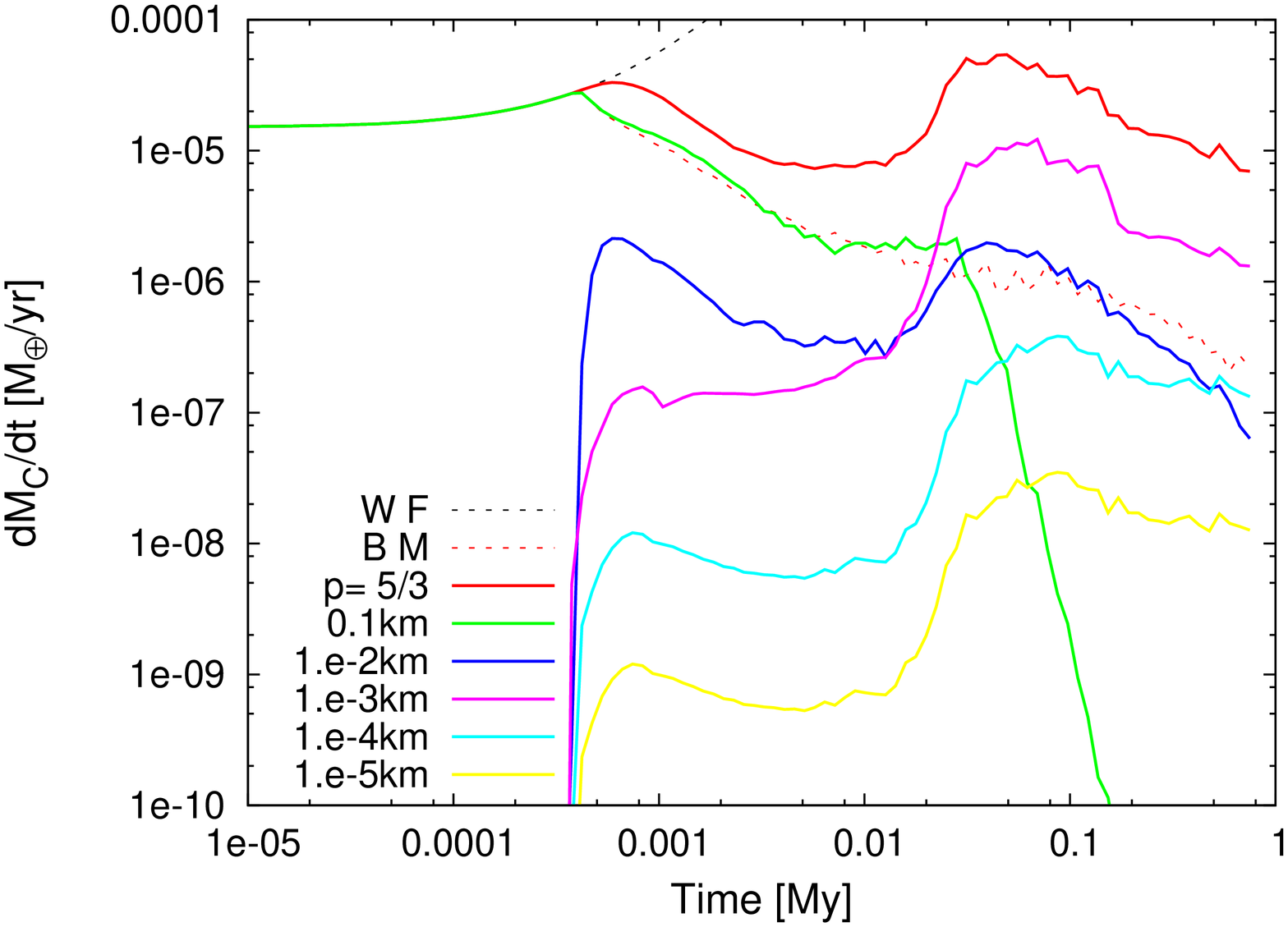}
  \includegraphics[angle=270, width= 0.45\textwidth]{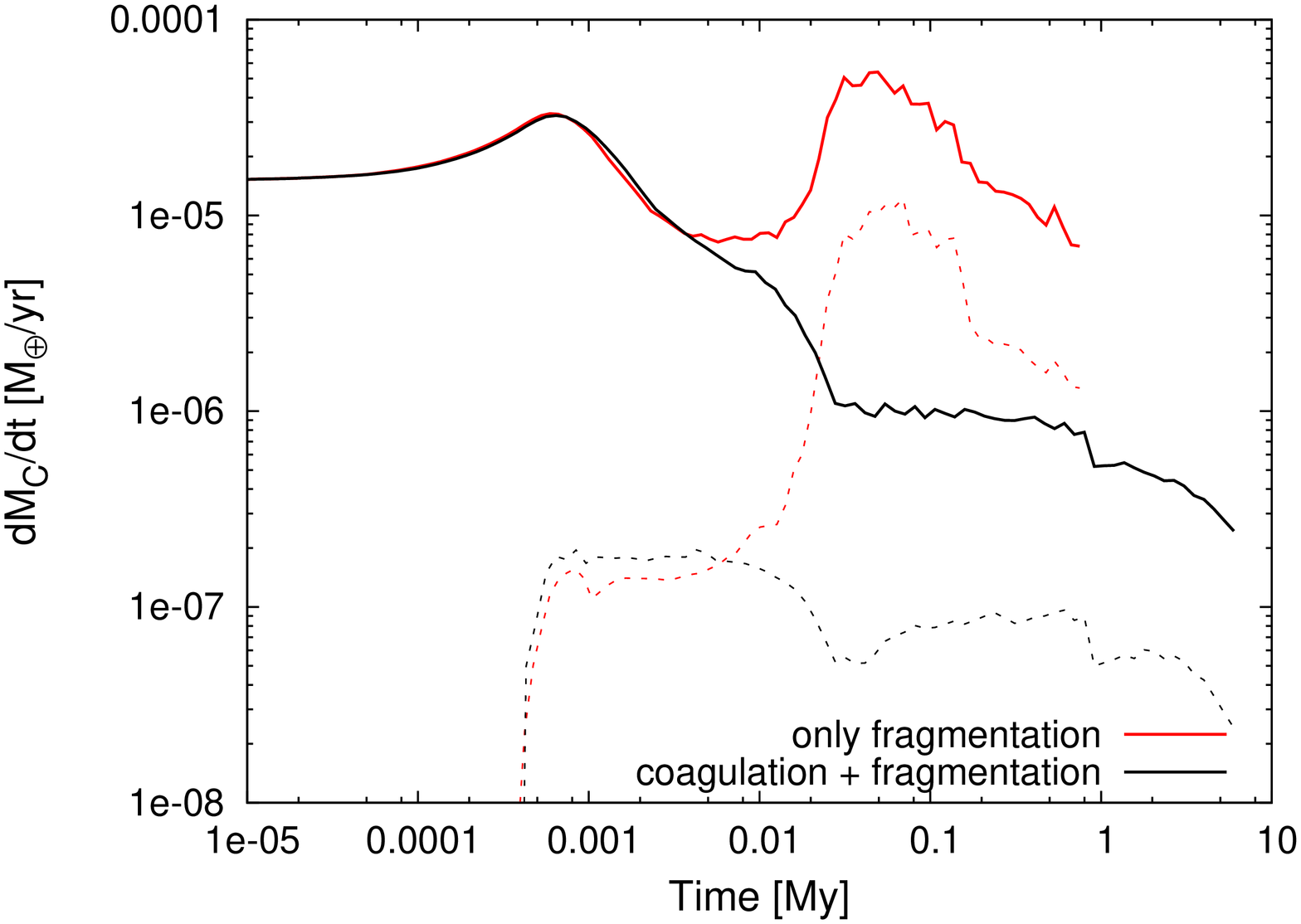}
  \caption{Top: planetesimal accretion rates for different planetesimal sizes, for the case of a disk 10 times more massive then the MMSN, in which $p= 5/3$ and $r_p^{max}= 0.1$~km, as function of time. Dashed black line represents the total planetesimal accretion rate for the case without
 planetesimal fragmentation (WF). Dashed red line represents
 the total planetesimal accretion rate for the case with
 planetesimal fragmentation in which $p$ is calculated with
 Eq.~(\ref{eq:expo_p}), our base model (BM). Solid red line represents
 the total planetesimal accretion rate for the case with
 planetesimal fragmentation with $p=5/3$. Bottom: time evolution of the total planetesimal accretion rates and the accretion rates of fragments of 1~m , for the case wherein only planetesimal fragmentation is considered (red lines) and for the case in which planetesimal coagulation and fragmentation are considered (black lines). The plot corresponds to a  disk 10 times more massive then the MMSN and for the case of $r_p^{max}= 0.1$~km. Color figure
 only available in the electronic version.} 
  \label{fig:tasas-rp10km}
\end{figure}

We also found that for small planetesimals, if most of the fragment mass is deposited in bigger
fragments the total planetesimal accretion rate becomes greater than the
corresponding to the case in which most of the fragment mass is deposited in smaller
fragments. However, for these cases the total planetesimal accretion is always lower than the corresponding for the case in which planetesimal fragmentation is not considered. We calculated again the simulations for the case of a disk 10 times more massive than the MMSN. However, for $r_p^{max}= 0.1, 1, 10$~km, the situation is different. For these cases, the accretion rate corresponding to $r_p^{max}$ quickly drops and the total accretion rate is dominated by fragments of $\sim r_p= 1$~m. However, collisions between these small fragments (and obviously with smaller fragments) are not disruptives but rather the outcome of a collision results in an effective accretion. So, in this scenario planetesimal coagulation is necessary. As example, in
Fig.~\ref{fig:tasas-rp10km} (top) we plot the planetesimal accretion rates
for the case of $r_p^{max}= 0.1$~km. We can see how the accretion rate of
planetesimals of radius 0.1~km significantly drops (green curve), and the total accretion rate is ultimately dominated by small planetesimals ($r_p \sim 1$~m, violet curve). The planetesimal accretion rates of fragments of $\sim r_p= 1$~m become significant, remaining high values of the total planetesimal accretion rates, but this effect is fictitious because small planetesimals coagulate forming larger bodies. For this case, we stopped the simulation at 0.75~My because at this time the core had reached a mass of $\sim 12~\textrm{M}_{\oplus}$. But again, these results could be fictitious. A planetesimals coagulation model is necessary in these cases.  

This effect does not happen for big planetesimals, wherein the accretion of such small fragments does not significantly contribute to the total accretion rate. In fact, for big planetesimals, the total accretion rate is always dominated by the accretion of planetesimals of radius $r_p^{max}$ and the fragments that have a significant contribution in the total accretion rate have radii greater than 0.1~km (collisions are disruptives for such fragments). 

However, in order to be more rigorous with these intuitive analysis, we recalculated the simulations for the cases of $r_p^{max}= 100$~km and $r_p^{max}= 0.1$~km, but incorporating also coagulation between planetesimals in the model. Following the Boulder code, we considered that the outcome of a collision results in accretion if the mass of the remnant is greater than the mass of the target. We considered that when a coagulation between planetesimals occurs, the target and projectile are removed from their corresponding radial bins, and a new object of mass $M_T + M_P$ is put in the radial bin corresponding to the target, i.e., we considered perfect accretion. We want to remark, that we have a very simple intention: we want to analyze if the coagulation between planetesimals modifies (or not) the results found in this section. So, for simplicity, we considered that the size grid that represents the continuous planetesimal size distribution was fixed. In spite of this, it is important to remark too that the computational costs are much greater.   

For the case of $r_p^{max}= 100$~km, we found identical results for both cases, considering only planetesimal fragmentation or considering planetesimal coagulation and fragmentation. As we argue before, this is because small fragments have a negligible contribution to the total planetesimal accretion rate (Fig.~\ref{fig:tasas-rp100km}).

However, for the case of $r_P^{max}= 0.1$~km, we found that the incorporation of planetesimal coagulation significantly modified the results. In Fig.~\ref{fig:tasas-rp10km} (bottom), we plot the time evolution of the total planetesimal accretion rate (solid lines) and the accretion rate of fragments of 1~m (dashed lines), for the case wherein only planetesimal fragmentation is considered (red lines) and for the case in which planetesimal coagulation and fragmentation are considered (black lines). For this last case, after 6~My of evolution the core achieved a mass of only $\sim 3~\textrm{M}_{\oplus}$. In both cases, the accretion of fragments of $~\sim 1$~m governed the total accretion rates, but the incorporation of planetesimal coagulation drastically diminishes the accretion of such fragments. It is clear that, when the accretion of very small fragments becomes important in the total planetesimal accretion rate, a full planetesimal collisional model (which includes coagulation and fragmentation) is needed. We will study in detail this topic developing a full planetesimal collisional model in a future work, incorporating an adaptative size grid to study more in detail the collisional evolution of the planetesimal population and starting from the planetesimal runaway growth. 

Finally, we want to remark that planetesimal coagulation did not quantitatively modify the results showed in Tab.~\ref{tab:general-results} for small values of $r_p^{max}$. For these cases, all the accretion rates for the different planetesimal sizes significantly diminish with time due to the fact that most of the mass loss in collisions is distributed below the size bin corresponding to $r_p= 1$~cm.

\subsection{About the accretion of small fragments}

Let we introduce a brief discussion about the accretion of small fragments (pebbles). Lambrechts \& Johansen (2012), found that pebbles are accreted extremely efficiently by embryos. The pebbles with the appropriate Stoke's number have a capture cross section as large as the Hill radius of the embryo, even if the embryo does not have a gaseous envelope. When they pass within the Hill radius, they spiral down to the embryo's physical radius due to gas drag. Lambrechts \& Johansen (2012) found that the pebbles accretion rates (in the Hill accretion regime) is given by, 
\begin{eqnarray}
\dot{M}_H= 2\textrm{R}_{\textrm{H}}\Sigma_pv_{H},
\end{eqnarray}
wherein $v_{H}= \Omega_k \textrm{R}_{\textrm{H}}$, being $\Omega_k$ the Keplerian frecuency. As the authors note, in the classical scenario of planetesimal accretion, planets do not accrete planetesimals from the Hill radius. Instead of this, planets accrete planetesimal from a fraction $\alpha^{1/2}$ of the Hill radius, with $\alpha= \textrm{R}_{\textrm{C}} / \textrm{R}_{\textrm{H}}$. 

In our work, we used the planetesimal accretion rates of Inaba et al. ( 2001) given by, 
\begin{eqnarray}
\dot{M}= 2\pi\Sigma_p\textrm{R}_{\textrm{H}}^2P_{\textrm{coll}}/P_{\textrm{orb}},
\end{eqnarray}

For small planetesimals, in the low velocity regime, the probability collision is given by, 
\begin{eqnarray}
P_{\textrm{coll}}= 11.3~\sqrt{\textrm{R}_{\textrm{C}} / \textrm{R}_{\textrm{H}}}. 
\end{eqnarray}
As we also consider the enhanced radius due to the planet's gaseous envelope $P_{\textrm{coll}}= 11.3~\sqrt{\tilde{\textrm{R}}_{\textrm{C}} / \textrm{R}_{\textrm{H}}}$. In term of the pebble accretion rate, our planetesimal accretion rate (for small fragments) is given by, 
\begin{eqnarray}
  \dot{M}= 5.65 \sqrt{\tilde{\textrm{R}}_{\textrm{C}} / \textrm{R}_{\textrm{H}}} ~ \dot{M}_H. 
  \label{eq:pebble_comparacion}
\end{eqnarray}          

\begin{figure}
  \centering
  \includegraphics[angle=270, width= 0.45\textwidth]{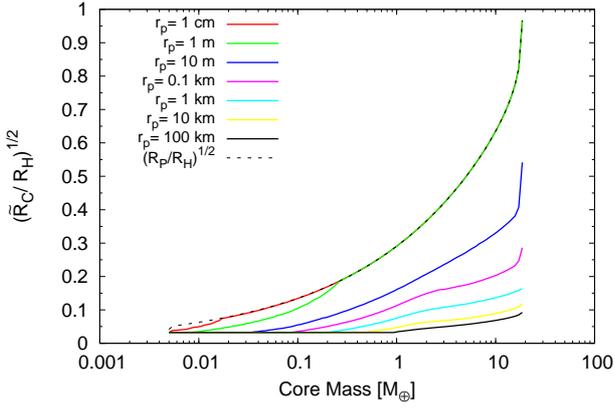}
  \caption{Factor $\sqrt{\tilde{\textrm{R}}_{\textrm{C}} / \textrm{R}_{\textrm{H}}}$ as function of the core mass for different planetesimal sizes. The plot corresponds to the case of a disk 10 times more masive than the MMSN and $r_p^{max}= 100$~km when planetesimal coagulation and fragmentation are considered. Black dashed line corresponds to the evolution of the term $(\textrm{R}_{\textrm{P}} / \textrm{R}_{\textrm{H}})^{1/2}$, being $\textrm{R}_{\textrm{P}}$ the radius of the planet. Color figure only available in the electronic version.}  
  \label{fig:pebbles}
\end{figure}

In Fig.~\ref{fig:pebbles} we plot the evolution of the term $\sqrt{\tilde{\textrm{R}}_{\textrm{C}} / \textrm{R}_{\textrm{H}}}$ as function of the core mass for different planetesimal sizes, for the case of a disk 10 times more masive than the MMSN and $r_p^{max}= 100$~km when planetesimal coagulation and fragmentation are considered. We can see that for small fragments ($r_p \le 1$~m) the evolution is almost the same. This is because for these small fragments the enhanced capture radius becomes the planet's radius\footnote{In our models the planet's radius is the radius of the envelope, which is the minimum between the accretion radius and the Hill radius, see Guilera et. al~(2010).} for small values of the core mass (black dashed lines). Due to the factor 5.65 in Eq.~(\ref{eq:pebble_comparacion}), our planetesimal accretion rates for $r_p \lesssim 1$~m are greater than the pebble accretion rates of Lambrechts \& Johansen (2012) when the planet's core mass become greater than $\sim 0.2~\textrm{M}_{\oplus}$. Despite these large accretion rates, these small fragments have a neglegible contribution in our models because the small values for the corresponding surface densities (see for example Fig.~\ref{fig:numero_cuerpos_0.1km}). It is important to note that the probability collision for smaller fragments ($r_p \lesssim 1$~m) is always the one correspondig to the low velocity regime.      

\subsection{Simultaneous formation of two embryos}   

Finally, we analyzed the in situ simultaneous formation of two embryos. We aimed study if the fragments generated by an outer embryo (which have an inward migration) favored the formation of an inner embryo. We only analyzed the case of a disk 10 times more massive than the MMSN, where initially all the solid mass of the system is deposited in planetesimals of $r_p=100$~km and for the case $p= 5/3$. We located the embryos at 5~AU and 6~AU. Both embryos have initially cores of $0.005~\textrm{M}_{\oplus}$ and envelopes of $\sim 10^{-13}~\textrm{M}_{\oplus}$. The simulation stopped at $\sim 2.5$~My. At this time, the embryo located at 5~AU achieved a total mass of $13.20~\textrm{M}_{\oplus}$ ($12.10~\textrm{M}_{\oplus}$ for the core and $1.10~\textrm{M}_{\oplus}$ for the envelope) while the embryo located at 6~AU achieved a total mass of $2.61~\textrm{M}_{\oplus}$ ($2.60~\textrm{M}_{\oplus}$ for the core and $0.01~\textrm{M}_{\oplus}$ for the envelope). The simulation stopped because the distance between embryos became lower than 3.5 mutual Hill radii. When two embryos are too close, their mutual gravitational perturbation may lead to encounters or collisions between them. 

\begin{figure}
  \centering
  \includegraphics[angle=270, width= 0.475\textwidth]{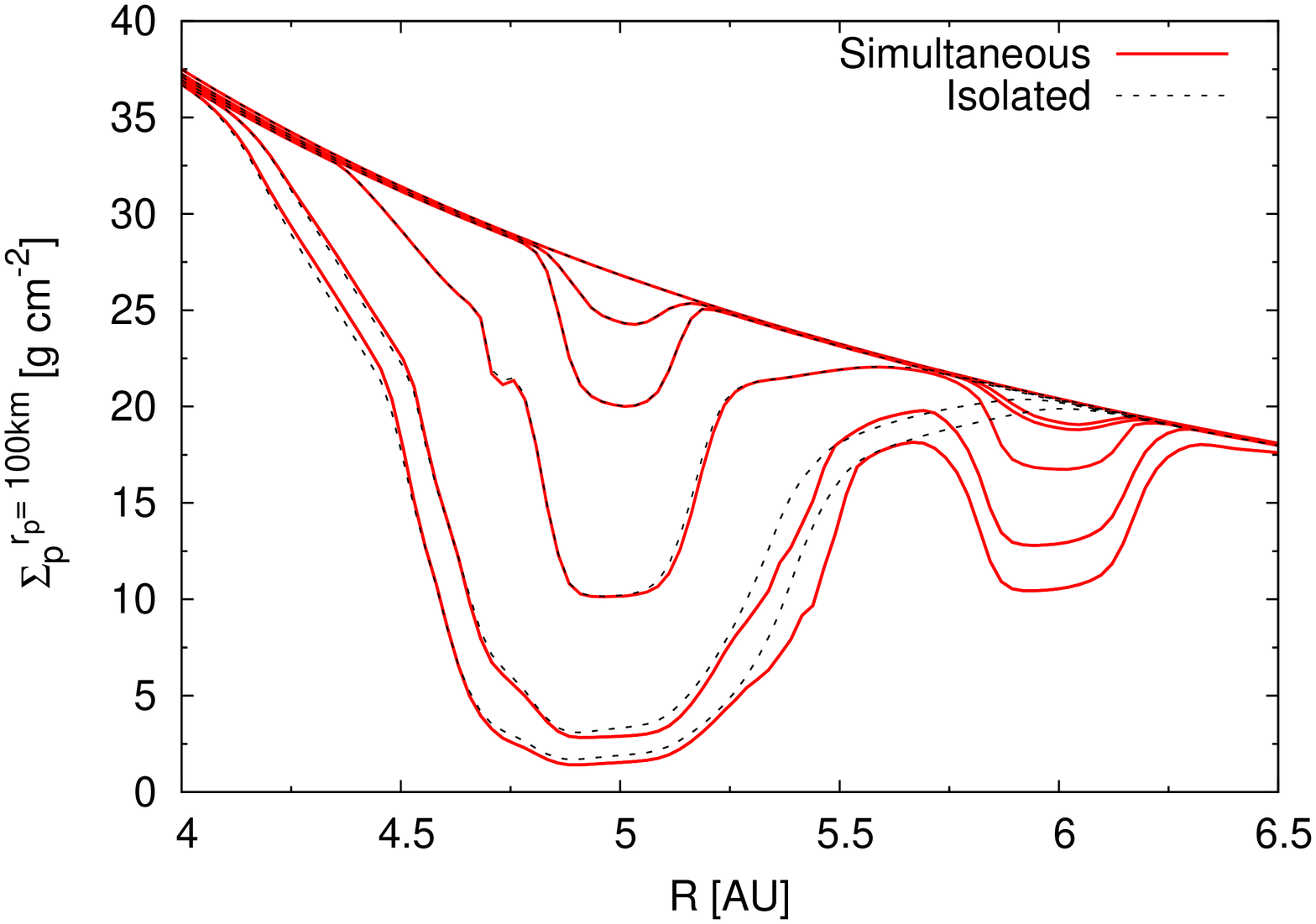}
  \includegraphics[angle=270, width= 0.475\textwidth]{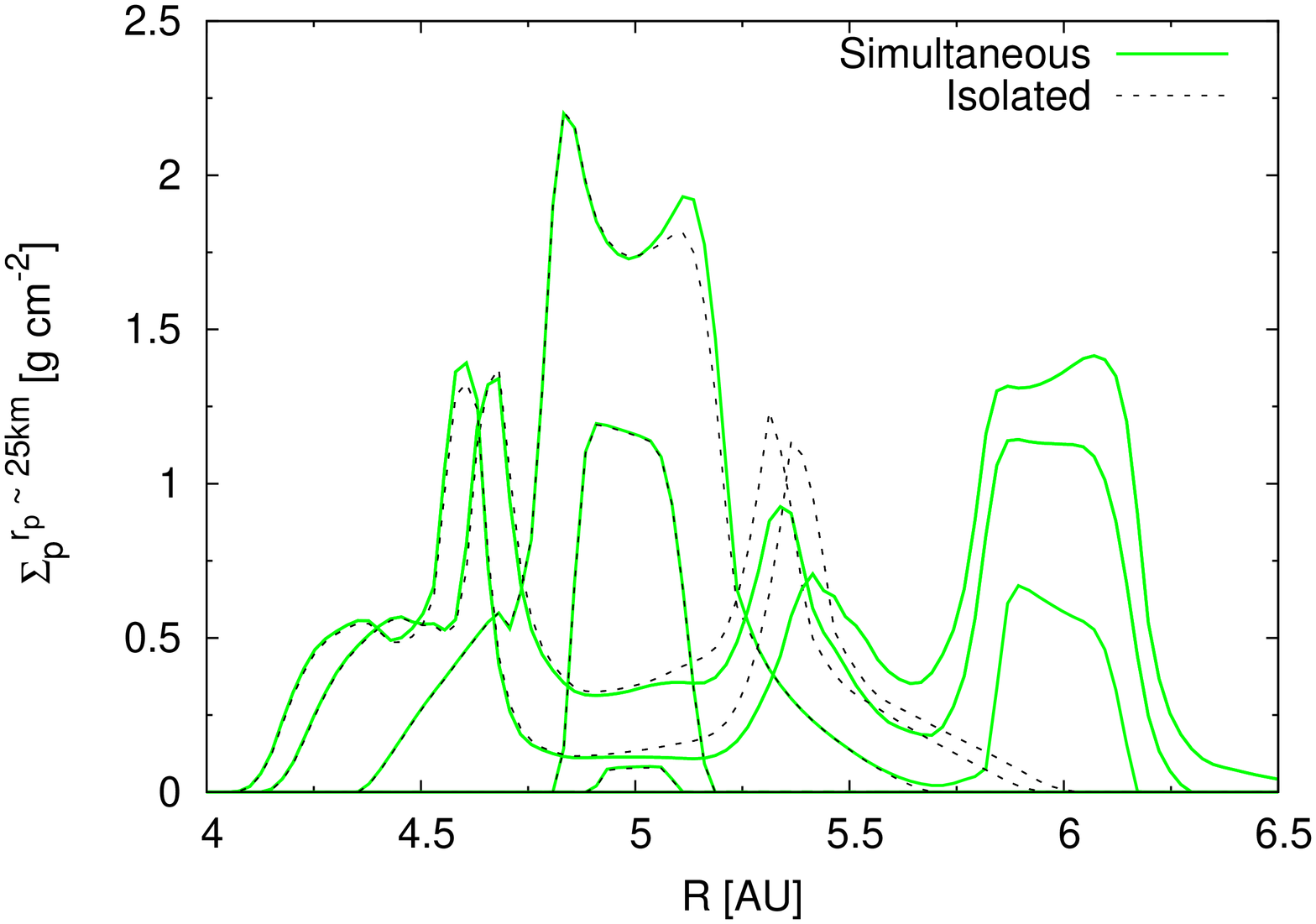}
  \caption{Top: Time evolution of the radial profiles of the surface density
 of planetesimals of $r_p= 100$~km when planetesimal fragmentation is
 considered with $p= 5/3$ and for a disk 10 times more massive than the
 MMSN. Black dashed lines correspond to the isolated formation of a
 planet located at 5~AU. Red solid lines correspond to the simultaneous
 formation of two planets located at 5~AU and 6~AU. The different lines
 correspond to different times: 0, 0.5, 1, 1.5, 2, $\sim 2.5$~My. As
 time advances, planetesimal surface density diminishes at embryos
 neighborhood. Bottom: same as top panel, but for fragments that most
 contribute in the total accretion rate, $r_p \sim 25$~km. In this case, the profiles increase around 5~AU for $t= 0.5, 1, 1.5$~My, but for $t= 2, \sim 2.5$~My profiles diminish because of accretion. The increments in the surface density around 6~AU correspond to the profiles at times 1.5, 2, $\sim 2.5$~My. Color figure only available in the electronic version.}     
 \label{fig:2P-perfiles}
\end{figure}

\begin{figure}
  \centering
  \includegraphics[angle=270, width= 0.475\textwidth]{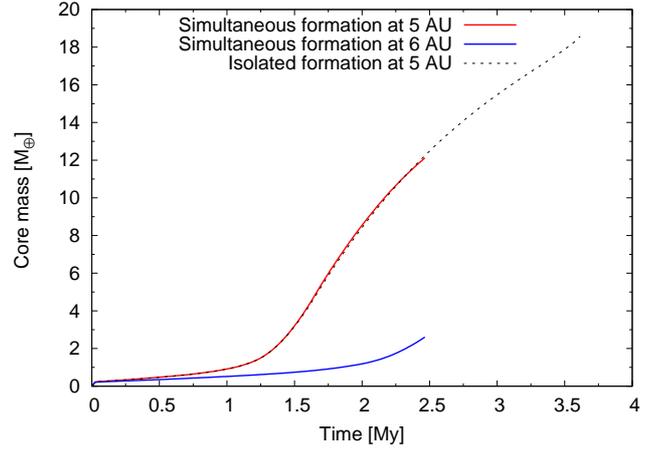}
  \caption{Core masses as function of time for the isolated formation of
 a planet located at 5~AU (black dashed line) and for the simultaneous
 formation of two planets located at 5~AU (red solid line) and 6~AU (blue solid
 line). Models correspond to $r_p^{max}= 100$~km for a disk 10 times
 more massive than the MMSN and planetesimal fragmentation is
 considered with $p=5/3$. The simultaneous formation stopped at
 $\sim 2.5$~My because the distance between planets became lower
 than 3.5 mutual Hill radii. Color figure only available in the
 electronic version.} 
 \label{fig:2P-masas}
\end{figure}

The time evolution of the radial profiles for the surface density of
planetesimals of $r_p= 100$~km and $r_p \sim 25$~km 
(which are the two sizes that
most contribute to the total planetesimal accretion rate, see
Fig.~\ref{fig:tasas-rp100km}, being the 25~km-sizes the fragments that
most contribute) for the embryo located at 5~AU are very similar
(Fig.~\ref{fig:2P-perfiles}). So, we do not expect significant
differences in the formation of this embryo. In fact, in
Fig.~\ref{fig:2P-masas} we plot the time evolution of the core mass for
both embryos comparing to the isolated embryo located at 5~AU. We do not
find differences in the formation of the embryo located at 5~AU between
isolated and the simultaneous formation with an outer embryo, at least
for the profile of the disk analyzed. 

%%%%%%%%%%%%%%%%%%% CONCLUSIONES %%%%%%%%%%%%%%%%%%%%%

\section{Conclusions}

Morbidelli et al.~(2009), employing the Boulder code, found that the present size frequency distribution of bodies in the asteroid belt can be reproduced starting with an initial population of big planetesimals between $100-1000$~km. However, Fortier et al.~(2009) demonstrated that the formation of massive cores able to achieve the critical mass to start the gaseous runaway phase, starting from big planetesimals, requires of several million years, even for massive disks.

In this work we studied the role of planetesimal fragmentation on giant
planet formation. We developed a model for planetesimal fragmentation
based in the Boulder code, and incorporated it in our model of giant planet formation (Guilera et al., 2010, 2011), so the planetesimal population evolved by planet accretion, migration and fragmentation. We numerically studied how planetesimal fragmentation modified the formation of an embryo located at 5~AU for a wide range of disk masses and planetesimal sizes.  

We considered that initially all the solid mass of the system is deposited
in planetesimals of radius $r_p^{max}$, and that the mass loss in
collisions is distributed by a power law mass distribution between the
biggest fragment and the minimum size bin. The exponent of such power
law mass distribution is calculated by the model. We found that this
exponent is generally greater than 2, so most of the mass is distributed
in the smaller fragments, below the size bin corresponding to 1~cm, which
is the minimum size that we considered for accretion. For this reason,
most of the mass produced by the collisions is lost and fragments did not
significantly contribute for the embryo growth.

We found that planetesimal fragmentation inhibits planet formation. Only for big planetesimals ($r_p^{max}= 100$~km), the mass of the embryo was greater than $1~\textrm{M}_{\oplus}$ for high mass disks, but for any case the core achieved the critical mass. However, if most of the mass loss in collisions is distributed in bigger fragments, i.e. the exponent of the power law mass distribution is lower than 2, like other works adopted (Kobayashi et al., 2011, Ormel \&
Kobayashi, 2012), we found that planetesimal fragmentation favored the relative rapid formation of a massive core (greater than $10~\textrm{M}_{\oplus}$). The accretion of fragments between $\sim100$~m and $\sim 25$~km increments the total
planetesimal accretion rate, but always this total accretion rate is governed by the accretion of planetesimals of $r_p= 100$~km. For this case, we also analyzed if the presence of an outer embryo modified the formation of an inner
one. So, we calculated the in situ simultaneous formation of two embryos
located at 5~AU and 6~AU. In particular, we wanted analyze if the
fragment migration produced by the outer embryo affected the
formation of the inner one. We did not find differences between
isolated and simultaneous formation for the embryo located at 5~AU. But
in this case, at $\sim 2.5$~My, separation between embryos became lower
than 3.5 mutual Hill radii, so simulation was stopped. Possible
mergers between massive embryos could lead to a rapid formation of
massive cores.

On the other hand, Weidenschilling (2011) showed that the present size
distribution observed in the asteroid belt can be also reproduced
starting from planetesimals of radius $\sim 0.1$~km. Kenyon \& Bromley
(2012) concluded that the size distribution of TNOs can be reproduced
starting from a massive disk composed by relative small planetesimals
($r_p \lesssim 10$~km). However, for such small planetesimals,
collisions between them quickly become highly catastrophic because of the
small values for the specific impact energy. So, targets are quickly
pulverized and the total surface density of solids drastically
drops. When the mass loss in collisions is distributed in smaller
fragments, like the Boulder code predicts, planet formation is
completely inhibited. When the mass loss in collisions is distributed in
bigger fragments, fragments of $\sim 1$~m ultimately govern the total
planetesimals accretion rate. However, for these small fragments
collisions between them result in accretion. So, for the case
wherein the mass loss in collisions is deposited in bigger fragments, we
repeated the simulations (for $r_p^{max}= 0.1, 100$~km) but
incorporating planetesimal coagulation. As we expected, planetesimal
coagulation did not modify the process of planetary formation for the
case of $r_p^{max}= 100$~km, due to the fact that small fragments have a
negligible contribution to the total planetesimal accretion rate. On the
other hand, planetesimal coagulation significantly modified the results
for $r_p^{max}= 0.1$~km, inhibiting the formation of a massive core. We will 
analyze in detail this case developing a full planetesimal collisional model in a future work.

%%%%%%%%%%% Agradecimientos %%%%%%%%%%%%%  
\begin{acknowledgements}
We thank the anonymous referee for their comments, which help us to improve the work. This work was supported by the PIP 112-200801-00712 grant from
 Consejo Nacional de Investigaciones Cient\'{\i}ficas y T\'ecnicas, and
 by G114 grant from Universidad Nacional de La Plata.  
\end{acknowledgements}

%%%%%%%%%%% Bibliografia %%%%%%%%%%%%%


\begin{thebibliography}{}

 \bibitem[]{} Adachi, I., Hayashi, C., \& Nakazawa, K.\ 1976, Progress of
	  Theoretical Physics, 56, 1756   

 \bibitem[]{} Alexander, R.~D., Clarke, C.~J., \& Pringle, J.~E.\ 2006,
	   \mnras, 369, 229  

 \bibitem[]{} Benz, W., \& Asphaug, E.\ 1999, \icarus, 142, 5 
  
 \bibitem[]{} Chambers, J.\ 2006, \icarus, 180, 496

 \bibitem[]{} Chambers, J.\ 2008, \icarus, 198, 256 

 \bibitem[]{} Fern{\'a}ndez, J.~A., Gallardo, T., \& Brunini, A.\ 2002,
	   \icarus, 159, 358  

\bibitem[]{} Fortier, A., Benvenuto, O.~G., \& Brunini, A.\ 2009, \aap, 500, 1249  

 \bibitem[]{} Fortier, A., Alibert, Y., Carron, F., Benz, W., \&
	   Dittkrist, K.-M.\ 2013, \aap, 549, A44  

 \bibitem[]{} Guilera, O.~M., Brunini, A., \& Benvenuto, O.~G.\ 2010,
	   \aap, 521, A50 


 \bibitem[]{} Guilera, O.~M., Fortier, A., Brunini, A., \& Benvenuto,
	   O.~G.\ 2011, \aap, 532, A142  
	   
 \bibitem[]{} Hasegawa, M., \& Nakazawa, K.\ 1990, \aap, 227, 619

 \bibitem[]{} Hayashi, C.\ 1981, Progress of  Theoretical Physics
	   Supplement, 70, 35  

 \bibitem[]{} Inaba, S., Tanaka, H., Nakazawa, K., Wetherill, G.~W., \&
	   Kokubo, E.\ 2001, \icarus, 149, 235 

 \bibitem[]{} Inaba, S., \& Ikoma, M.\ 2003, \aap, 410, 711 

 \bibitem[]{} Inaba, S., Wetherill,G. W., \& Ikoma, M.\ 2003, \icarus,
	   166, 46  

 \bibitem[]{} Kenyon, S.~J., \& Bromley, B.~C.\ 2012, \aj, 143, 63 

 \bibitem[]{} Kobayashi, H., \& Tanaka, H.\ 2010, \icarus, 206, 735

 \bibitem[]{} Kobayashi, H., Tanaka, H., Krivov, A.~V., \& Inaba, S.\
	   2010, \icarus, 209, 836  

 \bibitem[]{} Kobayashi, H., Tanaka, H., \& Krivov, A.~V.\ 2011, \apj,
	   738, 35  
	   
\bibitem[]{} Lambrechts, M., \& Johansen, A.\ 2012, \aap, 544, A32 

 \bibitem []{} Lissauer, J. J., \& Stevenson, D. J.\ 2007, Protostars and
	  Planets V, 591  

 \bibitem[]{} Morbidelli, A., Bottke, W.~F., Nesvorn{\'y}, D., \&
	   Levison, H.~F.\ 2009, \icarus, 204, 558  

 \bibitem[]{} Ohtsuki, K., Stewart, G.~R., \& Ida, S.\ 2002, \icarus,
	   155, 436  


 \bibitem[]{} Ormel, C.~W., \& Kobayashi, H.\ 2012, \apj, 747, 115  


 \bibitem[]{} Rafikov, R.~R.\ 2004, \aj, 128, 1348 
   
 \bibitem[]{} Thommes, E.~W., Duncan,  M.~J., \& Levison, H.~F.\ 2003,
	   \icarus, 161, 431 

 \bibitem[]{} Weidenschilling, S.~J.\ 2011, \icarus, 214, 671 
  

\end{thebibliography}
\end{document}